\documentclass[a4paper]{iopart}
\usepackage{bm,amssymb,epsfig,cite}
\newcommand{\be}{\begin{equation}}
\newcommand{\ee}{\end{equation}}
\newcommand{\bea}{\begin{eqnarray}}
\newcommand{\eea}{\end{eqnarray}}

\def\bra#1{|#1\rangle}

\def\g{\gamma}
\def\b{\beta}
\def\d{\delta}
\def\a{\alpha}
\def\e{\varepsilon}

\def\nn{\nonumber\\}

\def\r#1{(\ref{#1})}

\def\2t#1#2{\langle\tau_{#1}\tau_{#2}\rangle}

\def\nn{\nonumber\\}

\def\r#1{(\ref{#1})}

\def\i{{\rm i}}
\def\d{{\rm d}}
\def\e{{\rm e}}
\def\dps{\displaystyle}

\eqnobysec
\begin{document}

\title{Exact Spectral Gaps of the Asymmetric Exclusion Process with Open
  Boundaries}
\author{Jan de Gier$^1$ and Fabian H L Essler $^2$}
\address{$^1$ ARC Centre of Excellence for Mathematics and Statistics of
  Complex Systems, Department of Mathematics and Statistics, The University of
  Melbourne, 3010 VIC, Australia\\
$^2$ Rudolf Peierls Centre for Theoretical Physics, University
  of Oxford, 1 Keble Road, Oxford, OX1 3NP, United Kingdom}

\begin{abstract}
We derive the Bethe ansatz equations describing the complete spectrum
of the transition matrix of the partially asymmetric exclusion process
with the most general open boundary conditions. By analysing these
equations in detail for the cases of totally asymmetric and symmetric
diffusion, we calculate the finite-size scaling of the spectral gap,
which characterizes the approach to stationarity at large times. In
the totally asymmetric case we observe boundary induced crossovers
between massive, diffusive and KPZ scaling regimes. We further study
higher excitations, and demonstrate the absence of oscillatory
behaviour at large times on the ``coexistence line'', which separates
the massive low and high density phases. In the maximum current phase,
oscillations are present on the KPZ scale $t\propto L^{-3/2}$. While
independent of the boundary parameters, the spectral gap as well as
the oscillation frequency in the maximum current phase have different
values compared to the totally asymmetric exclusion process with
periodic boundary conditions. We discuss a possible interpretation of
our results in terms of an effective domain wall theory. 
\end{abstract}


\maketitle
The partially asymmetric simple exclusion process (PASEP) \cite{ASEP1,ASEP2}
is a model describing the asymmetric diffusion of hard-core particles
along a one-dimensional chain with $L$ sites. Over the last decade
it has become one of the most studied models of non-equilibrium
statistical mechanics, see \cite{Derrida98,Schuetz00} for recent
reviews. This is due to its close relationship to growth phenomena and the
KPZ equation \cite{KPZ}, its use as a microscopic model for driven
diffusive systems \cite{schmittmann} and shock formation
\cite{janowsky}, its applicability to molecular diffusion in zeolites
\cite{HahnKK96}, biopolymers \cite{biopolymer1,biopolymer2,biopolymer3}, traffic flow
\cite{ChowdSS00} and other one-dimensional complex systems \cite{Privman}. 

At large times the PASEP exhibits a relaxation towards a
non-equilibrium stationary state. An interesting feature of the PASEP
is the presence of boundary induced phase transitions \cite{Krug91}. 
In particular, in an open system with two boundaries at which
particles are injected and extracted with given rates, the bulk
behaviour in the stationary state is strongly dependent on the
injection and extraction rates. Over the last decade many stationary
state properties of the PASEP with open boundaries have been
determined exactly
\cite{Derrida98,Schuetz00,DEHP,gunter,sandow,EsslerR95,PASEPstat1,PASEPstat2}.  

On the other hand, much less is known about its dynamics. This is in
contrast to the PASEP on a ring for which exact results using Bethe's
ansatz have been available for a long time \cite{dhar,BAring1,BAring2}. 
For open boundaries there have been several studies of dynamical
properties based mainly on numerical and phenomenological methods 
\cite{numerics1,numerics2,numerics3,KSKS,DudzS00}. Very recently a real-space renormalization
group approach was introduced, which allows for the determination of
the dynamical exponents \cite{HaSt06}. 

In this work, elaborating on \cite{GierE05}, we employ Bethe's ansatz
to obtain exact results for the approach to stationarity at large
times in the PASEP with open boundaries. Upon varying the boundary
rates, we find crossovers in massive regions, with dynamic exponents
$z=0$, and between massive and scaling regions with diffusive ($z=2$)
and KPZ ($z=3/2$) behaviour. 

\begin{figure}[ht]
\centerline{
\begin{picture}(270,76)
\put(0,8){\epsfig{width=0.7\textwidth,file=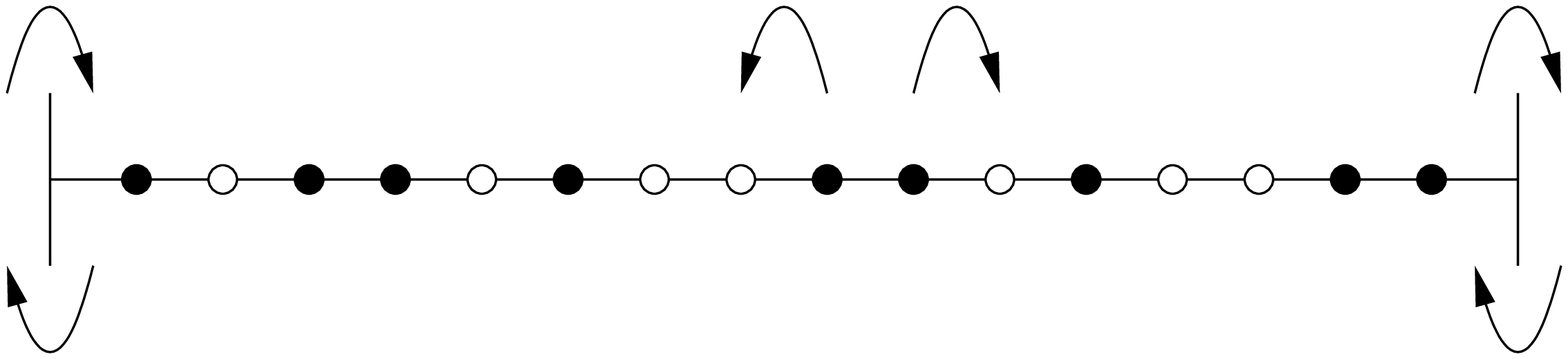}}
\put(4,0){$\gamma$}
\put(4,70){$\alpha$}
\put(249,0){$\delta$}
\put(249,70){$\beta$}
\put(129,71){$q$}
\put(157,71){$p$}
\end{picture}}
\caption{Transition rates for the partially asymmetric exclusion process.}
\label{fig:paseprules}
\end{figure}

The dynamical rules of the PASEP are as follows. At any given time
$t$ each site is either occupied by a particle or empty and the system
evolves subject to the following rules. In the bulk of the system
($i=2,\ldots,L-1$) a particle attempts to 
hop one site to the right with rate $p$ and one site to the left with
rate $q$. The hop is prohibited if the neighbouring site is
occupied. On the first and last sites these rules are modified. If
site $i=1$ is empty, a particle may  enter the system with rate
$\alpha$. If on the other hand site $1$ is occupied by a particle, the
latter will leave the system with rate $\gamma$. Similarly, at $i=L$
particles are injected and extracted with rates $\delta$ and $\beta$
respectively.  

With every site $i$ we associate a Boolean variable $\tau_i$, indicating whether a particle
is present ($\tau_i=1$) or not ($\tau_i=0$). Let $\bra0$ and $\bra1$
denote the standard basis vectors in $\mathbb{C}^2$. A state of the
system 
at time $t$ is then characterized by the probability distribution
\be
\bra{P(t)} = \sum_{\bm \tau} P(\bm{\tau}|t) \bra{\bm{\tau}},
\ee
where
\be
\bra{\bm{\tau}} = \bra{\tau_1,\ldots,\tau_L} = \bigotimes_{i=1}^{L} \bra{\tau_i}.
\ee
The time evolution of $\bra{P(t)}$ is governed by the aforementioned rules and as a result is subject to the
master equation    
\bea
\frac{{\rm d}}{{\rm d} t} \bra{P(t)} &=& M \bra{P(t)},
\label{eq:Markov}
\eea
where the PASEP transition matrix $M$ consists of two-body interactions only and is given by
\be
M = \sum_k I^{\otimes k-1} \otimes \widetilde{M} \otimes I^{\otimes L-k-1} + m_1 \otimes I^{\otimes L-1} + I^{\otimes L-1}\otimes m_L.
\ee
Here, $I$ is the identity matrix on $\mathbb{C}^2$ and $\widetilde{M}: \mathbb{C}^2\otimes \mathbb{C}^2 \rightarrow \mathbb{C}^2\otimes \mathbb{C}^2$ is given by
\be
\widetilde{M} = \left(\matrix{
0 & 0 & 0 & 0\cr
0 & -q & p & 0 \cr
0 & q & -p & 0 \cr
0 & 0 & 0 & 0}\right).
\ee
The boundary contributions $m_1$ and $m_L$ describe injection
(extraction) of particles with rates $\a$ and $\delta$ ($\g$ and $\b$) at
sites $1$ and $L$. In addition, as a tool to compute current
fluctuations \cite{DerridaDR04,Derrida05}, we introduce a fugacity
$\e^{\lambda}$ conjugate to the current on the first site. The
boundary contributions then are
\be
m_1=\left(\matrix{-\a&\g\e^{-\lambda} \cr \a \e^\lambda&-\g\cr}\right),\qquad 
m_L=\left(\matrix{-\delta&\b\cr \delta&-\b\cr}\right).
\label{h1l}
\ee
Strictly speaking, with the inclusion of $\lambda$ the matrix $M$ is
no longer a transition matrix of a stochastic process. In the following however,
we will still refer to $M$ as the transition matrix of the PASEP also
for nonzero values of $\lambda$.

At $\lambda=0$, the matrix $M$ has a unique stationary state
corresponding to eigenvalue zero. For positive rates, all
other eigenvalues of $M$ have non-positive real parts. The large time
behaviour of the PASEP is dominated by the eigenstates of $M$ with the
largest real parts of the corresponding eigenvalues. In the next
sections we will determine the eigenvalues of $M$ using Bethe's
ansatz. The latter reduces the problem of determining the spectrum of $M$
to solving a system of coupled polynomial equations of degree $3L-1$.
Using these equations, the spectrum of $M$ can be studied numerically
for very large $L$, and, as we will show, analytic results can be
obtained in the limit $L\rightarrow\infty$.

%%%%%%%%%%%%%%%%%%%%%%%%%%%%%%%%%%%%%%%%%%%%%%%%%%%%%%%%%%%%%%%%%
\section{Relation to the spin-1/2 Heisenberg XXZ chain 
with open boundaries\label{se:PASEP2XXZ}}
%%%%%%%%%%%%%%%%%%%%%%%%%%%%%%%%%%%%%%%%%%%%%%%%%%%%%%%%%%%%%%%%%
It is well known that the transition matrix $M$ is related to the
Hamiltonian $H$ of the open spin-1/2 XXZ quantum spin chain through a
similarity transformation  
\be
M = - \sqrt{pq}\, U_{\mu}^{-1} H_{\rm XXZ} U_{\mu},
\ee
where $U_\mu$ and $H_{\rm XXZ}$ are given by (see e.g. \cite{sandow,EsslerR95}), 
\bea
U_\mu &=& \bigotimes_{i=1}^L \left(\matrix{1 &0 \cr 0 &\mu Q^{j-1}}\right),\\  
H_{\rm XXZ} &=&\! -\frac12 \sum_{j=1}^{L-1} \left[
  \sigma_j^{\rm x}\sigma_{j+1}^{\rm x} + \sigma_j^{\rm
    y}\sigma_{j+1}^{\rm y} -\Delta \sigma_j^{\rm z}\sigma_{j+1}^{\rm
    z} + h (\sigma_{j+1}^{\rm z}-\sigma_j^{\rm z}) +\Delta \right]
\nonumber\\
 && {} +B_1 + B_L.
\label{eq:XXZham}
\eea
The parameters $\Delta$ and $h$, and the boundary terms $B_{1}$ and $B_L$
are related to the PASEP transition rates by
\bea
B_1 &=&\frac{1}{2\sqrt{pq}}\left(\alpha+\gamma+(\alpha-\gamma)
\sigma_1^{\rm z} - 2\alpha\mu \e^{\lambda} \sigma_1^{-} -
2\gamma\mu^{-1} \e^{-\lambda}
\sigma_1^{+}\right),\nn
B_L&=& \frac{1}{2\sqrt{pq}}\left(\beta+\delta-(\beta-\delta)
\sigma_L^{\rm z} - 2\delta\mu Q^{L-1}\sigma_L^{-} - 2\beta\mu^{-1} Q^{-L+1}
\sigma_L^{+}\right),\nn
\Delta&=& -\frac12(Q+Q^{-1}),\quad h=\frac12(Q-Q^{-1}),\quad
Q=\sqrt{\frac qp}.
\eea
Here $\sigma_j^{\rm x}$, $\sigma_j^{\rm y}$ and $\sigma_j^{\rm z}$ are
the usual Pauli matrices, $\sigma_j^{\pm} = (\sigma_j^{\rm x} \pm \i
\sigma_j^{\rm y})/2$, and $\mu$ is a free gauge parameter on which the
spectrum does not depend.

The expression in (\ref{eq:XXZham}) is the Hamiltonian of the
ferromagnetic $U_Q(SU(2))$ invariant quantum spin chain \cite{ps} with 
added boundary terms $B_1$ and $B_L$. We note that the boundary terms
contain non-diagonal contributions ($\sigma_1^\pm$, $\sigma_L^\pm$) with
$L$-dependent coefficients. In the absence of the boundary terms the
spectrum of the Hamiltonian is massive, i.e. there is a finite gap
between the absolute ground states and the lowest excited states.
As is shown below, the boundary terms lead to the emergence of several
different phases. Some of these phases exhibit a spectral gap in the
limit $L\to\infty$ in the sense that there is a finite gap in the real
part of the eigenvalue of the transition matrix. Other phases feature
gapless excited states. 

In order to make contact with the literature, we start by relating the
PASEP parameters to those used in previous analyses of the spin-1/2 XXZ
quantum spin chain with open boundaries. For the latter we employ
notations similar to \cite{magic05}, in which the XXZ Hamiltonian reads 
(up to a constant shift in energy)
\bea
H &=& -\frac12 \sum_{i=1}^{L-1}\left[\sigma^x_i \sigma^x_{i+1} + \sigma^y_i
\sigma^y_{i+1} - \cos \eta \left(\sigma^z_i \sigma^z_{i+1}-1\right)\right]
+{\cal B}_1+{\cal B}_L,\nn
{\cal B}_1&=&\frac{\sin\eta}{\cos\omega_-+\cos
  \delta_-}\Bigl[ \frac{\i}{2} (\cos \omega_--\cos\delta_-) 
\,\sigma_1^z \nn
&&\quad\qquad\qquad\qquad + \cos\theta_1 \sigma_1^x +\sin\theta_1 \sigma_1^y - \sin \omega_- \Bigr]\ ,\nn
 {\cal B}_L&=&\frac{\sin\eta}{\cos\omega_++\cos\delta_+}
\Bigl[-\frac{\i}{2} (\cos \omega_+ -\cos\delta_+)
\,\sigma_L^z \nn
&&\quad\qquad\qquad\qquad+ \cos\theta_2 \,\sigma_L^x + \sin \theta_2\, \sigma_L^y
 - \sin \omega_+ \Bigr].
\label{eq:XXZham2T}
\eea
Comparing this with the Hamiltonian (\ref{eq:XXZham}) arising from the
PASEP, we are led to the following identifications
\bea
&&Q=-\e^{\i\eta},\qquad \sqrt{\frac{\alpha}{\gamma}} = -\i
\e^{\i\omega_-},\qquad \sqrt{\frac{\beta}{\delta}} = -\i
\e^{\i\omega_+} \nonumber \\ 
&&\e^{\i \theta_1} = \sqrt{\frac{\alpha}{\gamma}} \mu\e^{\lambda},\qquad
\e^{\i\theta_2}= \sqrt{\frac{\delta}{\beta}} \mu Q^{L-1}.
\label{ident}
\eea
Furthermore, $\delta_\pm$ are determined via the equations
\be
-\sqrt{\frac{\alpha\gamma}{pq}} = \frac{\sin\eta}{\cos\omega_-+\cos\delta_-},\qquad
-\sqrt{\frac{\beta\delta}{pq}} = \frac{\sin\eta}{\cos\omega_++\cos\delta_+}.
\ee
As the spectrum of $H$ cannot depend on the gauge parameter
$\mu$, it follows from \r{ident} that the spectrum depends on
$\theta_1$ and $\theta_2$ only via $\theta_1-\theta_2$. 

Although it has been known for a long time that $H$ is integrable
\cite{GonzalesRuiz,Inami}, the off-diagonal boundary terms have so far
precluded a determination of the spectrum of $H$ by means of
e.g. the algebraic Bethe ansatz \cite{vladb}. However, recently a
breakthrough was achieved \cite{Cao03,Nepo02,NepoR03} in the case
where the parameters satisfy a certain constraint. In the above
notations this constraint takes the form 
\be
\cos(\theta_1-\theta_2) = \cos((2k+1)\eta+\omega_-+\omega_+).
\label{BAconstr}
\ee
Here $k$ is an integer in the interval $|k|\leq L/2$. 
In terms of the PASEP parameters the constraint reads
\be
(Q^{L+2k}-\e^{\lambda})(\alpha\beta\e^{\lambda} -
Q^{L-2k-2}\gamma\delta) =0.
\label{pasepconstr}
\ee
For given $k$ and $\lambda$ the constraint \r{pasepconstr} can be
satisfied in two distinct ways. Either one can choose $Q$ as a root of
the equation $Q^{L+2k}=e^\lambda$, or one can impose a relation on the
boundary and bulk parameters such that the second factor in
(\ref{pasepconstr}) vanishes. Curiously precisely under the latter
conditions, the DEHP algebra \cite{DEHP} for the PASEP fails to
produce the stationary state \cite{EsslerR95,ED}.  

Either choice results in a constraint on the allowed rates in the
PASEP. In particular this implies that it is not possible to calculate
current fluctuations for general values of the PASEP transition rates
from the Bethe ansatz equations presented below. 

However, in the case $\lambda=0$ it is possible to satisfy the
constraint (\ref{pasepconstr}) for arbitrary values of the PASEP
parameters by choosing $k=-L/2$.  As the condition \r{BAconstr} at
first sight appears not to have any special significance in terms of
the XXZ chain, it is rather remarkable that it can be identically
satisfied for the PASEP. 

%%%%%%%%%%%%%%%%%%%%%%%%%%%%%%%%%%%%%%%%%%%%%%%%%%
\subsection{Symmetries}
\label{se:sym}
%%%%%%%%%%%%%%%%%%%%%%%%%%%%%%%%%%%%%%%%%%%%%%%%%

The spectra of $M$ and $H$ are invariant under the particle-hole and left-right
symmetries of the PASEP: 

\begin{itemize}
\item Particle-hole symmetry
\be
\alpha \leftrightarrow \gamma,\qquad \beta \leftrightarrow
\delta,\qquad p \leftrightarrow q,\qquad \lambda \rightarrow -\lambda.
\ee

\item Left-right symmetry
\be
\alpha \leftrightarrow \delta,\qquad \beta \leftrightarrow
\gamma,\qquad p \leftrightarrow q,\qquad \lambda \rightarrow -\lambda.
\ee
\end{itemize}
Both particle-hole and left-right symmetries leave the two factors in
\r{pasepconstr} invariant individually. There is a third symmetry
which leaves \r{pasepconstr} invariant, but interchanges the two
factors 
\begin{itemize}
\item Gallavotti-Cohen symmetry \cite{Evans,GC} for the PASEP

\be
\e^{\lambda} \rightarrow \frac{\gamma\delta}{\alpha\beta}
Q^{2L-2} \e^{-\lambda}.
\label{eq:GC}
\ee
\end{itemize}
The combination of \r{eq:GC} with a redefinition of the gauge parameter,
\be
\mu \rightarrow \mu \e^{\lambda} Q^{-L+1}
\sqrt{\frac{\alpha\beta}{\gamma\delta}},
\ee
corresponds to the interchange $\theta_1 \leftrightarrow
\theta_2$. It is shown in the next section that if the constraint
\r{BAconstr} is satisfied, the spectrum of $H$ no longer depends
on the difference $\theta_1-\theta_2$. Hence \r{eq:GC} is not only a
symmetry of the constraint equation \r{BAconstr}, but also of the
spectrum of $H$. 

The Gallavotti-Cohen symmetry \r{eq:GC} implies
the fluctuation theorem \cite{Kurchan,LS} for the probability
$P_L(J,t)$ to observe a current $J$ on the first site at time $t$,
\be
\frac{P_L(-J,t)}{P_L(J,t)} \sim \left(\frac{\alpha\beta}{\gamma\delta}
Q^{-2L+2}\right)^{-Jt}\qquad (t\rightarrow\infty). 
\label{eq:FL}
\ee
We emphasize the $L$ dependence in \r{eq:GC} and \r{eq:FL}, which
disappears for symmetric hopping, $Q=1$ \cite{DDR}. It is further
clear from \r{eq:GC} that the current vanishes $J=0$ when the detailed
balance condition
\be
\frac{\alpha\beta}{\gamma\delta}
Q^{-2L+2} =1
\ee
is satisfied \cite{ED}. Precisely the same Gallavotti-Cohen symmetry as
in \r{eq:GC} was observed for the zero-range process (ZRP) with open
boundaries \cite{HRS}. This model is equivalent to the PASEP on an
infinite lattice but with a fixed number of $L+1$ particles and
particle dependent hopping rates. 

As a final remark, we note that it was realized in \cite{magic05} that
the constraint \r{BAconstr} has in fact an algebraic meaning and
corresponds to the non-semisimplicity of an underlying Temperley-Lieb
algebra, and implies the existence of indecomposable 
representations. Condition \r{BAconstr} can be interpreted as a
generalized root of unity condition for this algebra, and it implies
certain additional symmetries for the XXZ chain. It would be of interest
to understand the impact of these symmetries on the spectrum of the
PASEP with open boundaries\footnote{However, we have checked that for the TASEP
with open boundaries the spectrum consists of singlets only, confirming
\cite{GoliM04}, and hence is diagonalisable.}. 

%%%%%%%%%%%%%%%%%%%%%%%%%%%%%%%%%%%%%%%%%%%%%%%%%%%
\section{Bethe ansatz for the XXZ Hamiltonian} 
\label{se:BA}
%%%%%%%%%%%%%%%%%%%%%%%%%%%%%%%%%%%%%%%%%%%%%%%%%%%
The first Bethe ansatz results pertaining to the spectrum of
$H$ (\ref{eq:XXZham2T}) were reported in
\cite{Cao03,Nepo02}. Subsequently it was noted on the basis of
numerical  computations in \cite{NepoR03} and an analytical analysis
for a special case in \cite{GierP04}, that these initial results
seemed incomplete. Instead of one set of Bethe ansatz equations, one 
generally needs two. Further developments were reported in
\cite{Nepo05,YangNZ}. 

In order to simplify the following discussion we introduce the notations
\be
a_\pm = \frac{2\sin\eta\; \sin\omega_\pm}{\cos\omega_\pm + \cos\delta_\pm}.
\ee
When the constraint (\ref{BAconstr}) is satisfied for some integer
$k$, the eigenvalues of $H$ can be divided into two groups, $E_1(k)$
and $E_2(k)$
\bea E_1(k) &=& -a_- - a_+ - \sum_{j=1}^{L/2-1-k}
  \frac{2\sin^2\eta}{\cos 2u_j - \cos\eta},
  \label{eq:E2b-1}\\
E_2(k) &=& - \sum_{j=1}^{L/2+k}
  \frac{2\sin^2\eta}{\cos 2v_j - \cos\eta}.
  \label{eq:E2b-2}
\eea
Here the complex numbers $\{u_i\}$ and $\{v_i\}$ are solutions of
the coupled algebraic equations ($i=1,\ldots, L-1$)
\bea
w(u_i)^{2L} &=&
\frac{K_-(u_i-\omega_-)K_+(u_i-\omega_+)}
     {K_-(-u_i-\omega_-)K_+(-u_i-\omega_+)} \prod_{j=1 \atop
       {j\neq i}}^{L/2-1-k} \frac{S(u_i,u_j)}{S(-u_i,u_j)},
       \label{eq:BA2b-1}\\
w(v_i)^{2L} &=&
\frac{K_-(v_i)K_+(v_i)}
     {K_-(-v_i)K_+(-v_i)} \prod_{j=1 \atop {j\neq i}}^{L/2+k}
     \frac{S(v_i,v_j)}{S(-v_i,v_j)},
     \label{eq:BA2b-2}
\eea
where
\bea
w(u) &=& \frac{\sin(\eta/2+u_i)}{\sin(\eta/2-u_i)},\qquad S(u,v) =
\cos 2v - \cos(2\eta+2u),\nonumber\\
K_\pm(u) &=& \cos \delta_\pm + \cos(\eta+\omega_\pm
+2u).
%,\qquad \widetilde{K}_\pm(u) = \cos \delta_\pm + \cos(\gamma-\omega_\pm +2u).
\eea

We will now rewrite these equations in terms of the PASEP parameters
for which it is convenient to employ the notations
\be
z = -\e^{2\,\i u}, \qquad \zeta = -\e^{2\,\i v},
\ee
and
\bea
v_{\alpha,\gamma} &=& p-q-\alpha+\gamma,\\
\kappa^{\pm}_{\alpha,\gamma} &=& \frac{1}{2\alpha} \left(
v_{\alpha,\gamma} \pm \sqrt{v_{\alpha,\gamma}^2 +4\alpha\gamma}\right).
\eea
We then find that, \\
$\bullet$\; equations (\ref{eq:E2b-1}) and (\ref{eq:BA2b-1}) are
rewritten as
\bea
\sqrt{p q} E_1(k) = \alpha+\beta+\gamma+\delta+
\sum_{j=1}^{L/2-1-k}\frac{p\left(Q^2-1\right)^2
  z_j}{(Q-z_j)(Qz_j-1)},
\label{eq:pasep_en1}
\eea
\bea
\left[\frac{z_jQ-1}{Q-z_j}\right]^{2L} K_1(z_j) =\hskip -3pt\prod_{l\neq j}^{L/2-1-k}
\frac{z_jQ^2-z_l}{z_j-z_lQ^2} \frac{z_jz_lQ^2-1}
     {z_jz_l-Q^2},\ j=1\ldots L-1.\nn
\label{eq:pasep_eq1}
\eea
Here $K_1(z) = \tilde{K}_1(z,\alpha,\gamma) \tilde{K}_1(z,\beta,\delta)$ and
%
%\be
%\tilde{K}_1(z,\alpha,\gamma) =
%\frac{-\alpha z^2+Qz(q-p+\alpha-\gamma)+\gamma Q^2}{\gamma Q^2
%  z^2+Qz(q-p+\alpha-\gamma)-\alpha}. 
%\ee
\be
\tilde{K}_1(z,\alpha,\gamma) =
\frac{(z+Q\kappa^+_{\alpha,\gamma})
  (z+Q\kappa^-_{\alpha,\gamma})}{(Q\kappa^+_{\alpha,\gamma}z+1)
  (Q\kappa^-_{\alpha,\gamma}z+1)}. 
\ee
\bigskip\bigskip
$\bullet$\; the second set of equations, (\ref{eq:E2b-2}) and (\ref{eq:BA2b-2}), becomes
\bea
\sqrt{p q} E_2(k) = \sum_{j=1}^{L/2+k}\frac{p\left(Q^2-1\right)^2
  \zeta_j}{(Q-\zeta_j)(Q\zeta_j-1)},
\label{eq:pasep_en2}
\eea
\bea
\left[\frac{\zeta_jQ-1}{Q-\zeta_j}\right]^{2L} K_2(\zeta_j) =\prod_{l\neq j}^{L/2+k}
\frac{\zeta_jQ^2-\zeta_l}{\zeta_j-\zeta_lQ^2}
\frac{\zeta_j\zeta_lQ^2-1} {\zeta_j\zeta_l-Q^2},\ j=1\ldots L-1,\nn
\label{eq:pasep_eq2}
\eea
where $K_2(\zeta) = \tilde{K}_2(\zeta,\alpha,\gamma)
\tilde{K}_2(\zeta,\beta,\delta)$ and 
%
%\be
%\tilde{K}_2(\zeta,\alpha,\gamma) =
%\frac{\gamma \zeta^2+Q\zeta(q-p+\alpha-\gamma)-\alpha Q^2}{-\alpha Q^2
%  \zeta^2+Q\zeta(q-p+\alpha-\gamma)+\gamma}. 
%\ee
\be
\tilde{K}_2(z,\alpha,\gamma) =
\frac{(\kappa^+_{\alpha,\gamma}\zeta+Q)
  (\kappa^-_{\alpha,\gamma}\zeta+Q)}{(Q \zeta+ \kappa^+_{\alpha,\gamma})
  (Q \zeta+\kappa^-_{\alpha,\gamma})}. 
\ee

As we have mentioned before, in the case of the PASEP, and for
$\lambda=0$, the constraint (\ref{pasepconstr}) can be satisfied by
either considering symmetric hopping, $Q=1$, or by choosing $k=-L/2$.  

%%%%%%%%%%%%%%%%%%%%%%%%%%%%%%%%%%%%%%%%%%%%%%%%%%%%%%%
\section{Bethe ansatz equations for the ``generic'' PASEP}
%%%%%%%%%%%%%%%%%%%%%%%%%%%%%%%%%%%%%%%%%%%%%%%%%%%%%%%
Inspection of the second set of equations \r{eq:pasep_en2} for the
choice $k=-L/2$ reveals that there exists an isolated level with
energy ${\cal E}_0=0$. This is readily identified with the stationary state
of the PASEP.  Furthermore, \textit{all} other eigenvalues ${\cal E}$
of the transition matrix $M$ follow from the first set of equations
\r{eq:pasep_en1} and \r{eq:pasep_eq1}, and are given by  
\bea
{\cal E}= -\alpha-\beta-\gamma-\delta-\sum_{j=1}^{L-1}\frac{p\left(Q^2-1\right)^2
  z_j}{(Q-z_j)(Qz_j-1)},
\label{eq:pasep_en}
\eea
where the complex numbers $z_j$ satisfy the Bethe ansatz equations
\bea
\left[\frac{z_jQ-1}{Q-z_j}\right]^{2L} K(z_j) =\prod_{l\neq j}^{L-1}
\frac{z_jQ^2-z_l}{z_j-z_lQ^2} \frac{z_jz_lQ^2-1} {z_jz_l-Q^2},\
j=1\ldots L-1.\nn
\label{eq:pasep_eq}
\eea
Here $K(z) = \tilde{K}(z,\alpha,\gamma) \tilde{K}(z,\beta,\delta)$ and
%
%\be
%\tilde{K}(z,\alpha,\gamma) =
%\frac{-\alpha z^2+Qz(Q^2-1+\alpha-\gamma)+\gamma Q^2}{\gamma Q^2
%  z^2+Qz(Q^2-1+\alpha-\gamma)-\alpha}. 
%\ee
\be
\tilde{K}_1(z,\alpha,\gamma) =
\frac{(z+Q\kappa^+_{\alpha,\gamma})
  (z+Q\kappa^-_{\alpha,\gamma})}{(Q\kappa^+_{\alpha,\gamma}z+1)
  (Q\kappa^-_{\alpha,\gamma}z+1)}. 
\ee
In order to ease notations we set from now on, without loss of generality,
$p=1$ and hence $Q=\sqrt{q}$. 
%%%%%%%%%%%%%%%%%%%%%%%%%%%%%%%%%%%%%%%%%%%%%%%%
\section{Symmetric Exclusion Process (SEP)}
%%%%%%%%%%%%%%%%%%%%%%%%%%%%%%%%%%%%%%%%%%%%%%%%%
The limit of symmetric diffusion is quite special and we turn to it
next. We can obtain this limit either by taking $Q\to 1$ and leaving
$k$ unspecified in the equations \r{eq:pasep_en1}, \r{eq:pasep_eq1}
and \r{eq:pasep_en2}, \r{eq:pasep_eq2}, or by setting $k=-L/2$ and
studying the limit of the equations \r{eq:pasep_en}, \r{eq:pasep_eq} 
for the generic PASEP. The two procedures lead to the same results and
we will follow the second for the time being.

Taking the limit $Q\to 1$ in the equations (\ref{eq:pasep_en}),
(\ref{eq:pasep_eq}) for the general PASEP we observe that
\bea
&&\frac{zQ-1}{Q-z}\rightarrow  -1\ ,\qquad
\frac{z_jQ^2-z_l}{z_j-Q^2z_l} \rightarrow 1\ ,\nn
&&\frac{z_jz_lQ^2-1}{z_jz_l-Q^2} \rightarrow 1\ ,\qquad
\tilde{K}(z,\alpha,\gamma)\rightarrow -\frac{\alpha z+\gamma}{\gamma z+\alpha}.
\eea
We conclude that in this limit the spectral parameters $z$ must fulfill
\bea
\frac{\alpha z+\gamma}{\gamma z+\alpha}\frac{\beta z+\delta}{\delta
  z+\beta}=1\ .
\eea
It is easy to see that the only solutions to this equation are $z=\pm 1$.
Having established the leading behaviour of the roots of the Bethe
ansatz equations in the limit $Q\to 1$ we now parametrize
\be
z_j=\pm 1+\i\lambda_j(Q^2-1)\ ,
\label{param}
\ee
then substitute \r{param} back into the Bethe ansatz equations
\r{eq:pasep_eq}, \r{eq:pasep_en} and finally take the limit $Q\to 1$. 
Choosing the plus sign in \r{param}, we obtain with $c(x)=(x+2)/2x$
\bea
&\left(\frac{\lambda_j-\i/2}{\lambda_j+\i/2}\right)^{2L}&
%\frac{\lambda_j+i\frac{\alpha+\gamma+2}{2(\alpha+\gamma)}}
%{\lambda_j-i\frac{\alpha+\gamma+2}{2(\alpha+\gamma)}}
%\frac{\lambda_j+i\frac{\beta+\delta+2}{2(\beta+\delta)}}
%{\lambda_j-i\frac{\beta+\delta+2}{2(\beta+\delta)}}\nn
\left[\frac{\lambda_j+\i\, c(\alpha+\gamma)}{\lambda_j-\i\, c(\alpha+\gamma)}\right]
\left[\frac{\lambda_j+\i\, c(\beta+\delta)}{\lambda_j-\i\, c(\beta+\delta)}\right]\nn
&& \hspace{-1cm} =\prod_{l\neq  j}^{L-1}
\left[\frac{\lambda_j-\lambda_l-\i}{\lambda_j-\lambda_l+\i}\right]
\left[\frac{\lambda_j+\lambda_l-\i}{\lambda_j+\lambda_l+\i}\right], \quad
j=1,\ldots , L-1.
\label{baesymm}
\eea
\be
{\cal E}=-\alpha-\beta-\gamma-\delta-\sum_{j=1}^{L-1}\frac{1}{\lambda_j^2+1/4}.
\ee
On the other hand, choosing the minus sign in (\ref{param}) we arrive at
\bea
1=\prod_{l\neq  j}^{L-1}
\frac{\lambda_j-\lambda_l-\i}{\lambda_j-\lambda_l+\i}
\frac{\lambda_j+\lambda_l-\i}{\lambda_j+\lambda_l+\i}\ ,\qquad
j=1,\ldots , L-1.
\eea
\be
{\cal E}=-\alpha-\beta-\gamma-\delta\ .
\label{symmene2}
\ee
We observe that for this choice of sign in \r{param} we obtain only a
single energy level. 

For both choices there is a subtlety: we have implicitly assumed that
the $\lambda_j$'s remain finite when we take the limit $Q\to 1$ and
this need not be the case. A numerical analysis of the Bethe ansatz
equations \r{eq:pasep_eq} for $Q\approx 1$ shows that we have to allow
one or several $\lambda_j$'s to be strictly infinite, in which case
they are taken to drop out of (\ref{baesymm})
\footnote{We have verified this prescription by solving the Bethe
ansatz equations numerically for small systems in the vicinity of the
limit $Q\to 1$.}. This then
leads to the following set of Bethe ansatz equations, in which we only
allow solutions where all spectral parameters $\lambda_j$ are finite 
\bea
&\left(\frac{\lambda_j-\i/2}{\lambda_j+\i/2}\right)^{2L}&
\frac{\lambda_j+\i\, c(\alpha+\gamma)}{\lambda_j-\i\, c(\alpha+\gamma)}
\frac{\lambda_j+\i\, c(\beta+\delta)}{\lambda_j-\i\, c(\beta+\delta)}\nn
&&=\prod_{l\neq  j}^{N}
\frac{\lambda_j-\lambda_l-\i}{\lambda_j-\lambda_l+\i}
\frac{\lambda_j+\lambda_l-\i}{\lambda_j+\lambda_l+\i}\ ,
j=1,\ldots , N.
\label{baesymm2}
\eea
\be
{\cal E}=-\alpha-\beta-\gamma-\delta-\sum_{j=1}^{N}\frac{1}{\lambda_j^2+1/4}.
\label{baesymm2ene}
\ee
Here $N$ is allowed to take the values $1,2,\ldots L-1$.

Curiously, the limit of symmetric exclusion can be described by a
second set of Bethe ansatz equations. Taking the limit $Q\to 1$ of
(\ref{eq:pasep_en2}), (\ref{eq:pasep_eq2}) and leaving $k$ unspecified
we arrive at ${\cal E}=0$ or
\bea
&\left(\frac{\lambda_j-\i/2}{\lambda_j+\i/2}\right)^{2L}&
\frac{\lambda_j+\i\, d(\alpha+\gamma)}{\lambda_j-\i\, d(\alpha+\gamma)}
\frac{\lambda_j+\i\, d(\beta+\delta)}{\lambda_j-\i\, d(\beta+\delta)}\nn
&&=\prod_{l\neq  j}^{N}
\frac{\lambda_j-\lambda_l-\i}{\lambda_j-\lambda_l+\i}
\frac{\lambda_j+\lambda_l-\i}{\lambda_j+\lambda_l+\i}\ ,
j=1,\ldots , N,
\label{baesymm2b}
\eea
\be
{\cal E}=-\sum_{j=1}^{N}\frac{1}{\lambda_j^2+1/4}.
\label{baesymm2bene}
\ee
Here
\be
d(x)=c(-x)=\frac{x-2}{2x}.
\ee
Apart from a constant shift in energy the two sets of equations are
related by the simultaneous interchange $\alpha\leftrightarrow
-\gamma$ and $\beta\leftrightarrow -\delta$.
Importantly, equations \r{baesymm2b} coincide with the Bethe ansatz
equations derived in \cite{robin} by completely different means. Numerical
studies of small systems suggest that either set \r{baesymm} or
\r{baesymm2} gives the complete spectrum of the Hamiltonian.

Interestingly, the Bethe equations \r{baesymm2} and (minus) the
expression for the energy \r{baesymm2ene} are identical to the ones
for the open Heisenberg chain with boundary magnetic fields
\cite{openchain1,openchain2} \footnote{We note that for symmetric diffusion the
eigenvalues of the transition matrix are simply minus the corresponding eigenvalues
of the Heisenberg Hamiltonian.}
\bea
H=-2\sum_{j=1}^{L-1}\left[{\bf S}_j\cdot{\bf S}_{j+1}-\frac{1}{4}\right]
-\frac{1}{\xi_-}\left[S^z_1+\frac{1}{2}\right]
-\frac{1}{\xi_+}\left[S^z_L+\frac{1}{2}\right] ,
\eea
where $\xi_-=-1/(\alpha+\gamma)$ and
$\xi_+=-1/(\beta+\delta)$, if the reference state in the Bethe
ansatz is chosen to be the ferromagnetic state with all spins up. On
the other hand, the Bethe equations \r{baesymm2b} and (minus) the
expression for the energy \r{baesymm2bene} are obtained when the
reference state is chosen as the ferromagnetic state with all spins
down. This on the one hand shows the equivalence of the two sets of
Bethe equations and on the other hand establishes the fact, that the
spectrum of the SEP with open, particle number non-conserving boundary
conditions (which corresponds to the z-component of total spin $S^z$
not being a good quantum number in the spin-chain language) is
identical to the spectrum of the open Heisenberg chain with boundary 
magnetic fields, for which $S^z$ is a conserved quantity. This
spectral equivalence is more easily established by  means of a
similarity transformation, see section 6.5.1 \cite{Schuetz00} and
\cite{ADHR} for the case at hand and \cite{Baj}
for a general discussion on spectral equivalences for conserving and
non-conserving spin chains. 

The first excited state for the SEP occurs in the sector $N=1$ of
\r{baesymm2b}. The solution to the Bethe ansatz equations for large
$L$ is
\bea
\lambda_1 &=&\frac{L}{\pi}-\frac{1}{\pi}\left[d(\alpha+\gamma)+d(\beta+\delta)
\right]-\frac{\pi}{12L}\nn
&-&\frac{\pi}{6L^2}\left[d(\alpha+\gamma)-2d^3(\alpha+\gamma)
+d(\beta+\delta)-2d^3(\beta+\delta)\right]+\ldots
\eea
The corresponding eigenvalue of the transition matrix scales like
$L^{-2}$ with a coefficient that is independent of the boundary rates
\be
{\cal E}_1(L)=-\frac{\pi^2}{L^2}+{\cal O}(L^{-3}).
\ee
We conclude that for the SEP the large time relaxational behaviour is
diffusive and universal.
%%%%%%%%%%%%%%%%%%%%%%%%%%%%%%%%%%%%%%%%%%%%%%%%%%%%%%%
\section{Totally Asymmetric Exclusion Process (TASEP)}
%%%%%%%%%%%%%%%%%%%%%%%%%%%%%%%%%%%%%%%%%%%%%%%%%%%%%%%
We now turn to the limit of totally asymmetric exclusion $Q=0$ and
set $\gamma=\delta=0$ in order to simplify the analysis.
%%%%%%%%%%%%%%%%%%%%%%%%%%%%%%%%%%%%%%%%%%%%%
\subsection{Stationary State Phase Diagram}
\begin{figure}[ht]
\begin{center}
\epsfxsize=0.6\textwidth
\epsfbox{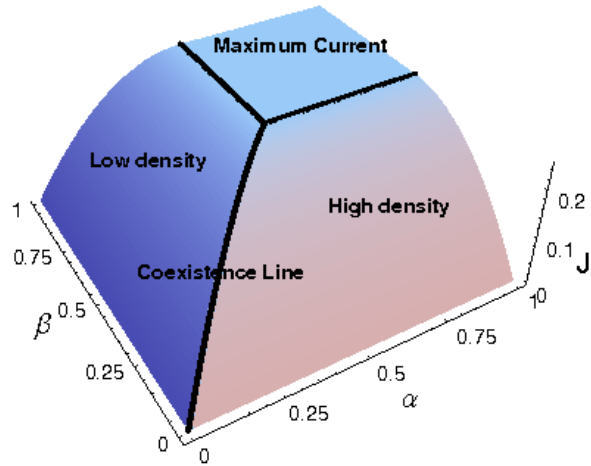}
\caption{Stationary phase diagram determined by the current of the
  TASEP.}
\label{fig:statphase}
\end{center}
\end{figure}

%%%%%%%%%%%%%%%%%%%%%%%%%%%%%%%%%%%%%%%%%%%%%
The stationary state phase diagram for the TASEP was determined in
\cite{DDM,gunter,DEHP} and features four distinct phases, see
Figure~\ref{fig:statphase}:

\begin{itemize}
\item[1.] A low density phase for $\alpha<1/2$, $\alpha<\beta$.

In the low-density phase the current in the thermodynamic limit is
equal to $J=\alpha(1-\alpha)$ and the density profile in the bulk is
constant $\rho=\alpha$.
\item[2.] A high density phase for $\beta<1/2$, $\beta<\alpha$. 

Here the current in the thermodynamic limit is
equal to $J=\beta(1-\beta)$ and the density profile in the bulk is
constant $\rho=1-\beta$.
\item[3.] A coexistence line at $\beta=\alpha<1/2$.

On the coexistence line the current is equal to $J=\alpha(1-\alpha)$,
but the density profile increases linearly in the bulk
$\rho(x)=\alpha+(1-2\alpha)x$. 

\item[4.] A maximal current phase at $\alpha,\beta>1/2$.

This phase is characterized by the current taking the maximal possible
value $J=1/4$ and the bulk density being constant and equal to
$\rho=1/2$. 
\end{itemize}
In \cite{gunter} a further subdivision of the low and high density
phases was proposed on the basis of differences in the behaviour of
the density profile in the vicinity of the boundaries. In the
high-density phase this distinction corresponds to the parameter
regimes $\alpha<1/2$ and $\alpha>1/2$, respectively.

%%%%%%%%%%%%%%%%%%%%%%%%%%%%%%%%%%%%%%%%%%%%%%%%%%%%%%%
\subsection{Analysis of the Bethe ansatz equations}
%%%%%%%%%%%%%%%%%%%%%%%%%%%%%%%%%%%%%%%%%%%%%%%%%%%%%%%
\label{se:BAanalysis}

In order to determine the exact value of the spectral gap we will now
analyse (\ref{eq:pasep_en}) and (\ref{eq:pasep_eq}) in the limit
$L\rightarrow\infty$. 

After rescaling $z \rightarrow Q z$ and setting $\gamma=\delta=0$, the
$Q\rightarrow 0$ limit of equations \r{eq:pasep_en} and
\r{eq:pasep_eq} reads 
\bea
&& {\cal E} = -\alpha-\beta - \sum_{l=1}^{L-1} \frac{z_l}{z_l-1},
\label{eq:tasep_en}\\
&&\left(\frac{(z_j-1)^2}{z_j}\right)^{L} = \left(z_j +
a\right) \left(z_j + b\right) \prod_{l\neq j}^{L-1} \left(
z_j-z_l^{-1}\right).
\label{eq:tasepBAE}
\eea
In order to ease notations in what follows, we introduce
\bea
g_{\rm }(z) &=& \ln \left( \frac{z}{(z-1)^2}\right),\label{eq:g}\\
g_{\rm b}(z) &=& \ln \left(\frac{z}{1-z^2}\right) +
\ln\left(z+a\right) + \ln\left(z+b\right),\label{eq:gb}
\eea
where
\be
a =\frac{1}{\alpha}-1,\qquad b=\frac{1}{\beta}-1\ .
\ee
The central object of our analysis is the ``counting function''
\cite{yaya69,deVegaW85,book},  
\be
\i Y_L(z) = g_{\rm}(z) + \frac{1}{L} g_{\rm b}(z)
+ \frac{1}{L} \sum_{l=1}^{L-1} K(z_l,z),
\label{eq:logtasepBAE}
\ee
where $K(w,z)$ is given by
\be
K(w,z) = -\ln w + \ln(1-w z).
\ee
Using the counting function, the Bethe ansatz equations
\r{eq:tasepBAE} can be cast in logarithmic form as
\be
Y_L(z_j) = \frac{2\pi}{L} I_j\ ,\qquad j=1,\ldots,L-1.
\label{eq:Z=I}
\ee
Here $I_j$ are integer numbers. The eigenvalues \r{eq:tasep_en} of the
transition matrix can be expressed in terms of the counting function as
\be
{\cal E} = -\alpha-\beta - L \lim_{z\rightarrow 1} \left( \i\, Y_L'(z) -
  g'(z) - \frac{1}{L} g'_{\rm b}(z)\right).
\label{eq:EinY}
\ee

\begin{figure}[ht]
\centerline{
\begin{picture}(350,100)
\put(0,0){\epsfig{width=150pt,file=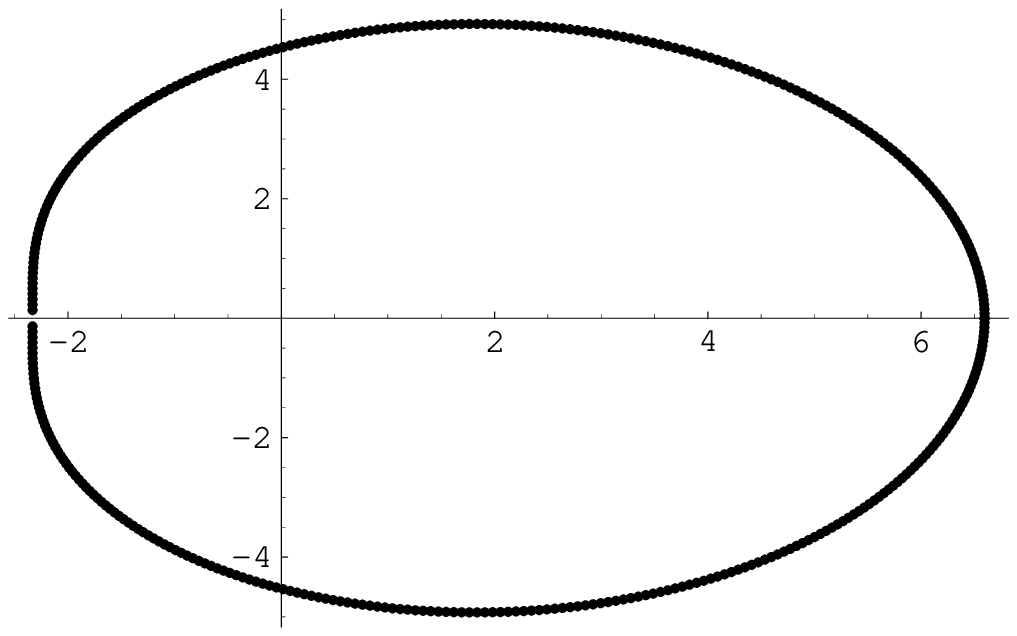}}
\put(0,0){(a)}
\put(200,0){\epsfig{width=150pt,file=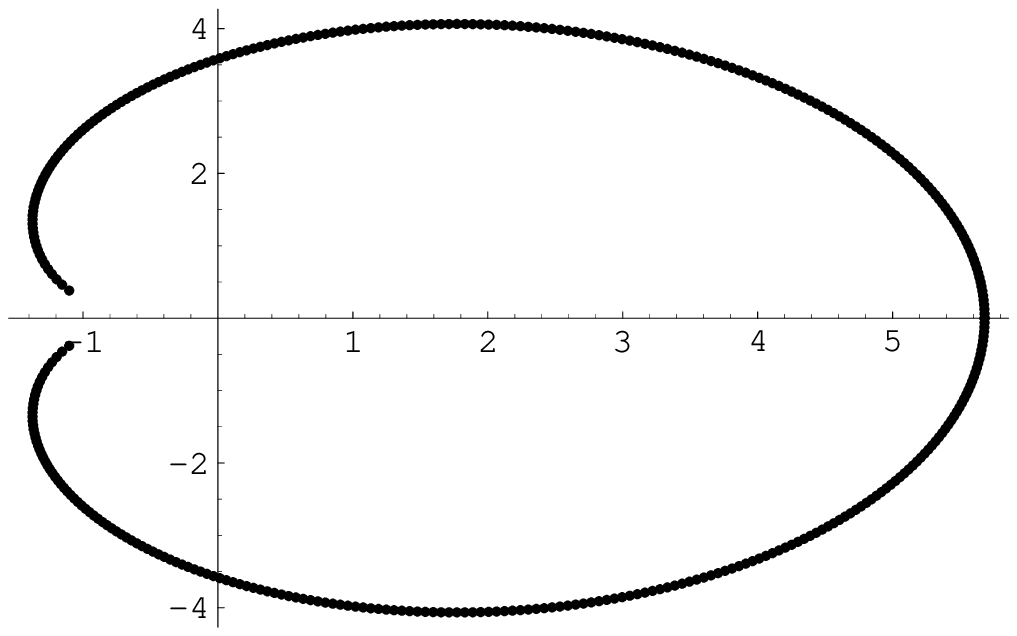}}
\put(200,0){(b)}
\end{picture}}
\caption{Reciprocal root distributions for (a) $\alpha=\beta=0.3$ and
  (b) $\alpha=\beta=0.7$, both with $L=2n=398$} 
\label{fig:roots}
\end{figure}

Each set of integers $\{I_j|j=1,\ldots, L-1\}$ in (\ref{eq:Z=I})
specifies a particular excited state. In order to determine which set
corresponds to the first excited state, we have calculated the
eigenvalues of the transition matrix numerically for small systems of
up to $L=14$ sites for many different values of $\alpha$ and $\beta$. 
By comparing these with the results of a numerical solution of the
Bethe ansatz equations, we arrive at the conclusion that
the first excited state always corresponds to the same set of integers
\be
I_j = -L/2+j\quad {\rm for}\quad j=1,\ldots,L-1.
\label{eq:Idef}
\ee
The corresponding roots lie on a simple curve in the complex plane,
which approaches a closed contour as $L\rightarrow \infty$. The
latter fact is more easily appreciated by considering the locus of
reciprocal roots $z_j^{-1}$ rather than the locus of roots $z_j$.
In Figure~\ref{fig:roots} we present results for $\alpha=\beta=0.3$ and
$\alpha=\beta=0.7$ respectively. The limiting shape of the curve is
that of the cardioid, which can be seen as follows. Assuming that the
last term in (\ref{eq:logtasepBAE}) is approximately constant as
$L\rightarrow\infty$ and using \r{eq:Z=I}, (\ref{eq:Idef}) we find that
\be
\exp(-g(z_j)) = (z_j^{1/2}-z_j^{-1/2})^2 = \chi \e^{-2\pi\i j/L} \qquad
(L\rightarrow\infty).
\label{eq:approx}
\ee
Here $\chi=\chi(\alpha,\beta)$ is some constant. Parametrising
$z=\rho\e^{\i\theta}$ and multiplying \r{eq:approx} by its complex
conjugate, we conclude that the roots lie on the curve defined by
\be
\rho^2 - (\chi+2\cos\theta)\rho+1=0,
\ee
the defining equation of the cardioid.

The shape of the locus of inverse roots depends on the rates $\alpha$
and $\beta$. For example, in the case $\alpha=\beta=0.7$ shown in
Figure~\ref{fig:roots}(b), a cusp is seen to develop at the intercept
of the curve with the negative real axis (which occurs at $z=-1$ in
the limit $L\to\infty$). In terms of the parameter $\chi$
characterizing the 
cardioid the cusp develops at $\chi=4$. In contrast, no cusp
occurs for $\alpha=\beta=0.3$ as shown in Figure~\ref{fig:roots}(a).\footnote{This
  corresponds to having $\chi>4$.}  
We will see that this difference in the shape of the loci of inverse
roots is reflected in a profoundly different finite-size scaling
behaviour of the corresponding spectral gaps.

\begin{figure}[ht]
\begin{center}
\begin{picture}(145,115)
\put(0,0){\epsfig{width=145pt,file=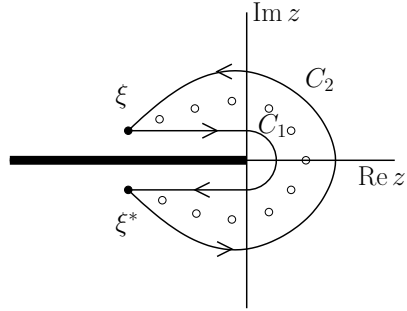}}
\end{picture}
\end{center}
\caption{Sketch of the contour of integration $C$ in \r{eq:sum2int}. The
  open dots correspond to the roots $z_j$ and $\xi$ is chosen close to
  $z_{L-1}$ and avoiding poles of $\cot(LY_L(z)/2)$.} 
\label{fig:contour}
\end{figure}

In order to compute the exact large $L$ asymptotics of the spectral
gap, we derive an integro-differential equation for the counting
function $Y_L(z)$ in the limit $L\rightarrow\infty$. As a simple
consequence of the residue theorem we can write 
\be
\frac1L \sum_{j=1}^{L-1} f(z_j) = \oint_{C_1+C_2} 
\frac{dz}{4\pi\i}\ f(z)
Y_L'(z) \cot\left(\frac12 L Y_L(z)\right),
\label{eq:sum2int}
\ee
where $C=C_1+C_2$ is a contour enclosing all the roots $z_j$, $C_1$
being the ``interior'' and $C_2$ the ``exterior'' part, see
Figure~\ref{fig:contour}. The contours $C_1$ and $C_2$ intersect in
appropriately chosen points $\xi$ and $\xi^*$. It is convenient to fix
the end points $\xi$ and $\xi^*$ by the requirement
\be
Y_L(\xi^*) = -\pi +\frac{\pi}{L},\qquad Y_L(\xi) = \pi -\frac{\pi}{L}.
\label{eq:xidef}
\ee
Using the fact that integration from $\xi^*$ to $\xi$ over the contour
formed by the roots is equal to half that over $C_2 - C_1$ we find,
\bea
\i\,Y_L(z) &=& g(z) + \frac{1}{L} g_{\rm b}(z) +\frac{1}{2\pi}
\int_{\xi^*}^{\xi} K(w,z) Y'_L(w) \d w \nonumber \\ 
&& \hspace{-1cm}+ \frac{1}{2\pi} \int_{C_1} \frac{K(w,z)Y'_L(w)}{1-\e^{-\i L
Y_L(w)}}\, \d w + \frac{1}{2\pi} \int_{C_2} \frac{K(w,z)Y'_L(w)}{
\e^{\i L Y_L(w)}-1}\,\d w,
\label{eq:intY}
\eea
where we have chosen the branch cut of $K(w,z)$ to lie along the
negative real axis. 

Our strategy is to solve the integro-differential equation \r{eq:intY}
by iteration and then use the result to obtain the eigenvalue
of the transition matrix from equation \r{eq:EinY}.

%%%%%%%%%%%%%%%%%%%%%%%%%%%%%%%%%%%%%
\section{Low and High Density Phases}
%%%%%%%%%%%%%%%%%%%%%%%%%%%%%%%%%%%%%
\label{se:LandH}
In the low and high density phases the locations of the end points $\xi$
and $\xi^*$ are such that a straightforward expansion of \r{eq:intY} in
inverse powers of $L$ is possible (see e.g. \cite{Olver,PovoPH03}
and \ref{app:abel}). 
The result is
\bea
\i\,Y_L(z) &=& g(z) + \frac{1}{L} g_{\rm b}(z) +\frac{1}{2\pi}
\int_{\xi^*}^{\xi} K(w,z) Y'_L(w) \d w \nonumber\\
&& {} + \frac{\pi}{12L^2} \left(
\frac{K'(\xi^*,z)}{Y'_L(\xi^*)} - \frac{K'(\xi,z)}{Y'_L(\xi)}\right) +
     \mathcal{O}(L^{-4})\nonumber\\
&=& g(z) + \frac{1}{L} g_{\rm b}(z) +\frac{1}{2\pi}
\int_{z_{\rm c}^-}^{z_{\rm c}^+} K(w,z) Y'_L(w) \d w \nonumber\\
&&{} + \frac1{2\pi} \int_{\xi^*}^{z_{\rm c}^-} K(w,z) Y'_L(w) \d w + \frac1{2\pi} \int_{z_{\rm c}^-}^{\xi} K(w,z) Y'_L(w) \d w \nonumber\\
&& {} + \frac{\pi}{12L^2} \left(
\frac{K'(\xi^*,z)}{Y'_L(\xi^*)} - \frac{K'(\xi,z)}{Y'_L(\xi)}\right) +
     \mathcal{O}(L^{-4})\ ,
\label{eq:intY_M}
\eea
where the derivatives of $K$ are with respect to the first
argument. We note that here we have implicitly assumed that
$Y'_L(\xi)$ is nonzero and of order ${\cal O}(L^0)$. In order to find the
eigenvalue of the transition matrix \r{eq:EinY} up to second order in
inverse powers of $L$, we will need to solve \r{eq:intY_M}
perturbatively to third order. Substituting the expansions
\be
Y_L(z) = \sum_{n=0}^\infty L^{-n} y_n(z),\qquad \xi = z_{\rm c} +
\sum_{n=1}^\infty L^{-n} (\delta_n + \i \eta_n),
\label{eq:expansion}
\ee
back into (\ref{eq:intY_M}) yields a hierarchy of equations for the
functions $y_n(z)$ of the type 
\be
y_n(z) = g_n(z) + \frac{1}{2\pi \i} \int_{z_{\rm c}^-}^{z_{\rm c}^+}
K(w,z) y'_n(w) \,\d w,
\label{eq:yn}
\ee
where $z_{\rm c}^{\pm} = z_{\rm c} \pm \i 0$. The integral is along
the closed contour following the locus of the roots. The first few
driving terms $g_n(z)$ are given by
\be
\renewcommand{\arraystretch}{1.2}
\begin{array}{r@{\hspace{2pt}}c@{\hspace{3pt}}l}
g_0(z) &=& -\i g(z), \\
g_1(z) &=& -\i g_{\rm b}(z) + \kappa_1 + \lambda_1 \tilde{K}(z_{\rm c},z),
\\
g_2(z) &=& \kappa_2 + \lambda_2 \tilde{K}(z_{\rm c},z) + \mu_2 K'(z_{\rm
  c},z), \\
g_3(z) &=& \kappa_3 + \lambda_3 \tilde{K}(z_{\rm c},z) + \mu_3 K'(z_{\rm
  c},z) + \nu_3 K''(z_{\rm c},z).
\end{array}
\label{eq:gs}
\ee
We recall that the functions $g$ and $g_{\rm b}$ are defined in \r{eq:g} and
\r{eq:gb} and $\tilde{K}(z_{\rm c},z) = \ln (z-z_{\rm c}^{-1})$. The
coefficients $\kappa_n$, $\lambda_n$, $\mu_n$ and $\nu_n$ are given in
terms of $\delta_n$, $\eta_n$ defined by \r{eq:expansion} as well as
derivatives of $y_n$ evaluated at $z_{\rm c}$. Explicit expressions are
presented in \ref{ap:Mcoef}. 

These coefficients as well as $z_{\rm c}$ are determined
self-consistently by solving \r{eq:yn} and then imposing the boundary
conditions \r{eq:xidef}. 
 
%%%%%%%%%%%%%%%%%%%%%%%%%%%%%%%%%%%%%%%%%%%%%%%%%%%
\subsection{Small values of $\alpha$ and $\beta$ }
%%%%%%%%%%%%%%%%%%%%%%%%%%%%%%%%%%%%%%%%%%%%%%%%%%%
%

When $\alpha$ and $\beta$ are small we assume that the singularities
in $g_{\rm b}(z)$, i.e. the points $-a=1-1/\alpha$ and $-b=1-1/\beta$,
lie outside the contour of integration. From the 
distribution of the reciprocal roots, Figure~\ref{fig:roots}(a), we
further infer that for small values of $\alpha$ and $\beta$ the roots
lie inside the unit circle. In particular we assume that $z_{\rm
  c}\neq -1$ and that the points $\pm 1$ lie outside the contour of
integration. 

We now proceed to solve \r{eq:yn} and then verify a posteriori that
the above assumptions hold.

The equation \r{eq:yn} for $n=0$ is solved by the simple ansatz
\be
y_0(z) = \kappa_0 + g_0(z)\ .
\ee
Substituting the ansatz into the integro-differential equation
\r{eq:yn} for $n=0$ we find
that 
\bea
\kappa_0 &=& - \frac{1}{2\pi} \int_{z_{\rm c}^-}^{z_{\rm c}^+} (-\ln w + \ln
(1-wz)) \left(\frac1w -\frac{2}{w-1}\right)\,\d w \nonumber\\
&=& -\i \left( -\ln(-z_{\rm c}) + 2 \ln (1-z_{\rm c})\right).
\eea
This in turn implies that the zeroth order term in the expansion of
the counting function is given by
\be
y_{0}(z) = -\i \ln \left[ - \frac{z}{z_{\rm c}}
\left(\frac{1-z_{\rm c}}{1-z}\right)^2 \right].
\label{eq:Z0sol}
\ee
In order to derive this result we have made use of the following
simple but useful identity ($C$ denotes the contour of integration
from $z_{\rm c}^-$ to $z_{\rm c}^+$)
\be
\frac{1}{2\pi\i} \int_{z_{\rm c}^-}^{z_{\rm c}^+} \frac{\ln w}{w+x}
\,\d w = \left\{ 
\begin{array}{ll}
\ln(1+z_{\rm c}/x) \quad & {\rm if} -x\ {\rm outside}\ C,\\
\ln(-x-z_{\rm c}) &  {\rm if} -x\ {\rm inside}\ C.
\end{array}\right.
\label{eq:logint}
\ee
The integro-differential equations for $n=1,2,3$ are solved in an
analogous manner, with the results
\bea
y_1(z) &=& - \i \ln\left[ -\frac{z}{z_{\rm c}} \frac{1-z_{\rm c}^{2}}{1-z^{2}}
\left(\frac{z_{\rm c}-z_{\rm c}^{-1}}{z-z_{\rm c}^{-1}}\right)^{-\i\lambda_1}
\frac{z+a}{z_{\rm c}+a} \frac{z+b}{z_{\rm 
    c}+b}\right]\nonumber\\
&& +\kappa_1-\i\ln \left(ab (-z_{\rm c})^{-\i\lambda_1}\right)
\label{eq:y1sol},\\
y_2(z) &=& \lambda_2 \ln \left(\frac{z_{\rm c}-z_{\rm
    c}^{-1}}{z-z_{\rm c}^{-1}}\right) + \frac{\mu_2}{z_{\rm c}^2}
    \left( \frac{1}{z-z_{\rm c}^{-1}} - \frac{1}{z_{\rm c}-z_{\rm
    c}^{-1}} \right)\nonumber\\
&& {}+\kappa_2 - \lambda_2 \ln (-z_{\rm c}) - \frac{\mu_2}{z_{\rm c}},
\label{eq:y2sol}\\
y_3(z) &=& \lambda_3 \ln \left(\frac{z_{\rm c}-z_{\rm
    c}^{-1}}{z-z_{\rm c}^{-1}}\right) + \left(\frac{\mu_3}{z_{\rm
    c}^2} - \frac{\nu_3}{z_{\rm c}^3}\right) 
    \left( \frac{1}{z-z_{\rm c}^{-1}} - \frac{1}{z_{\rm c}-z_{\rm
    c}^{-1}} \right)\nonumber\\
&& {} -\frac{\nu_3}{z_{\rm c}^3} \left( \frac{z}{(z-z_{\rm c}^{-1})^2} - \frac{z_{\rm c}}{(z_{\rm c}-z_{\rm
    c}^{-1})^2} \right) +\kappa_3 - \lambda_3 \ln (-z_{\rm c}) \nonumber\\
    &&{}- \left( \frac{\mu_3}{z_{\rm c}}-\frac{\nu_3}{z_{\rm c}^2}\right).
\label{eq:y3sol}
\eea
The coefficients $\kappa_n$, $\lambda_n$, $\mu_n$ and $\nu_n$ are
given in terms of $\delta_n$, $\eta_n$ and derivatives of $y_n$ at
$z_{\rm c}$, see \ref{ap:Mcoef} and we now proceed to determine them.
Substituting the expansions \r{eq:expansion} into equation
\r{eq:xidef}, which fixes the endpoints $\xi$ and $\xi^*$, we obtain a
hierarchy of conditions for $y_n(z_{\rm c})$, e.g.
\bea
Y_L(\xi)&=&y_0(\xi)+\frac{1}{L}y_1(\xi)+\frac{1}{L^2}y_2(\xi)+\ldots\nn
&=&y_0(z_{\rm c})+\frac{1}{L}\left[y_1(z_{\rm c})+y_0'(z_{\rm c})(\delta_1+i\eta_1)\right]+\ldots\nn
&=&\pi-\frac{\pi}{L}\ .
\eea
Solving these order by order we obtain
\be
\renewcommand{\arraystretch}{1.2}
\begin{array}{r@{\hspace{2pt}}c@{\hspace{3pt}}l}
\lambda_1 &=& 2\i,\\
\lambda_3 &=& \mu_2 = \lambda_2 = \kappa_1=0,\\
\nu_3 &=& \dps z_{\rm c} \mu_3 = z_{\rm c} (z_{\rm c}-1)^2 \kappa_2 =
-\i \pi^2 z_{\rm c}^2 \left(\frac{z_{\rm c}-1}{z_{\rm c}+1}\right)^2, 
\end{array}
\ee
with $\kappa_3$ undetermined. Furthermore, the intersect of
the solution curve with the negative real axis in the limit
$L\rightarrow \infty$ is found to be
\be
z_{\rm c} = -\frac{1}{\sqrt{a b}}.
\ee
Having determined the counting function, we are now in a position to
evaluate the corresponding eigenvalue of the transition matrix from
equation \r{eq:EinY}
\bea
{\cal E}_1(L) &=& -\alpha-\beta - \frac{2z_{\rm c}}{1-z_{\rm c}} -
\frac{\pi^2}{z_{\rm c}-z_{\rm c}^{-1}} \frac1{L^2} +
\mathcal{O}\left(L^{-3}\right) \nonumber\\ 
&=& -\alpha-\beta + \frac{2}{1+\sqrt{a b}}  - \frac{\pi^2}{\sqrt{a
    b}-1/\sqrt{a b}} \frac{1}{L^2} + \mathcal{O}\left(L^{-3}\right).
\label{eq:energyMI}
\eea
We conclude that to leading order in $L$, the eigenvalue of the
transition matrix with the second largest real part is a nonzero
constant. This implies an exponentially fast relaxation to the
stationary state at large times. We note that due to the symmetry of
the root distribution corresponding to \r{eq:Idef} under complex
conjugation ${\cal E}_1$ is in fact real.
The domain of validity of \r{eq:energyMI} is determined by the initial
assumption that $-a$ and $-b$ lie outside the contour of integration,
i.e.  
\be
-a < z_{\rm c} \quad\textrm{and}\quad -b < z_{\rm c}.
\ee
The parameter regime in $\alpha$ and $\beta$ in which \r{eq:energyMI}
is valid is therefore bounded by the curves
\be
\renewcommand{\arraystretch}{2.6}
\begin{array}{r@{\hspace{2pt}}c@{\hspace{3pt}}l}
\beta_{\rm c} &=& \dps
\left[1+\left(\frac{\alpha}{1-\alpha}\right)^{1/3}\right]^{-1}\ ,\quad
0\leq \alpha < \frac12,\\
\alpha_{\rm c} &=& \dps
\left[1+\left(\frac{\beta}{1-\beta}\right)^{1/3}\right]^{-1}\ ,\quad 0\leq
\beta < \frac12.
\end{array}
\label{eq:curves}
\ee
%
%%%%%%%%%%%%%%%%%%%%%%%%%%%%%%%%
\subsubsection{Coexistence Line}
%%%%%%%%%%%%%%%%%%%%%%%%%%%%%%%%
On the line $\beta=\alpha$, the leading term in \r{eq:energyMI} vanishes
and the eigenvalue with the largest real part different from zero is
therefore
\be
{\cal E}_1(L) = -\frac{\pi^2\alpha(1-\alpha)}{1-2\alpha}\frac{1}{L^2} +
\mathcal{O}(L^{-3}). 
\label{eq:energyCL}
\ee
This equation holds for fixed $\alpha<1/2$ and $L\to\infty$.
We note that there is a divergence for $\alpha\to 1/2$,
signaling the presence of a phase transition.
The inverse proportionality of the eigenvalue (\ref{eq:energyCL}) 
to the square of the system size implies a dynamic exponent
$z=2$, which suggests that the dominant relaxation at large
times is governed by diffusive behaviour. 
As shown in \cite{KSKS,DudzS00} the diffusive nature of the relaxational
mechanism can in fact be understood in terms of the unbiased random
walk behaviour of a shock (domain wall between a low and high density
region). Our results \r{eq:energyMI} and \r{eq:energyCL} for the phase
domain given by \r{eq:curves} and the coexistence line agree with the
relaxation time calculated in the framework of a domain wall theory
(DWT) model in \cite{DudzS00}. As we will show next, this is in
contrast to the massive phases beyond the phase boundaries
\r{eq:curves}, where the exact result will differ from the DWT
prediction. In Section~\ref{se:summ} we present a modified DWT
framework that allows us to understand the phase boundaries
\r{eq:curves}.  

%%%%%%%%%%%%%%%%%%%%%%%%%%%%%%
\subsection{Massive phase II:}
%%%%%%%%%%%%%%%%%%%%%%%%%%%%%%
In this section we will treat the case $-b > z_{\rm c}$ with $-a < z_{\rm
c}$ as before. The case $-a > z_{\rm c}$ with $-b < z_{\rm c}$ is
obtained by the interchange $\alpha\leftrightarrow \beta$ in the
relevant formulas. 

We need to solve the same integral equation \r{eq:yn} as before; in
particular the driving terms $g_n$ defined in \r{eq:gs} remain
unchanged. However, when iterating the driving term $g_1(z)$ there is
an extra contribution because the pole at $-b$ now lies inside the
integration contour, see \r{eq:logint}. As a result the solution for
$y_1(z)$ is now of the form
\bea
y_1(z) &=&  - \i \ln\left[ -\frac{z}{z_{\rm c}} \frac{1-z_{\rm c}^{2}}{1-z^{2}}
\left[\frac{z_{\rm c}-z_{\rm c}^{-1}}{z-z_{\rm c}^{-1}}\right]^{-\i\lambda_1}
\frac{z+a}{z_{\rm c}+a} \frac{z+b}{z_{\rm 
    c}+b}\frac{z+1/b}{z_{\rm c}+1/b}\right] \nonumber\\
&& +\kappa_1 -\i\ln \left(a (-z_{\rm c})^{-\i\lambda_1}\right).
\eea
The forms of the solutions for $y_n$ for $n\geq 2$ remain
unchanged. The coefficients of the various terms are again fixed by
imposing the boundary condition (\ref{eq:xidef}), with the result
\be
\renewcommand{\arraystretch}{1.2}
\begin{array}{r@{\hspace{2pt}}c@{\hspace{3pt}}l}
\lambda_1 &=& 3\i,\\
\lambda_3 &=& \mu_2 = \lambda_2 = \kappa_1=0,\\
\nu_3 &=& \dps z_{\rm c} \mu_3 = \frac32 z_{\rm c} (z_{\rm c}-1)^2 \kappa_2 =
-4 \i \pi^2 z_{\rm c}^2 \left(\frac{z_{\rm c}-1}{z_{\rm c}+1}\right)^2,
\end{array}
\ee
and $\kappa_3$ again remains undetermined. We note that the difference in
the functional form of $y_1(z)$ compared to \r{eq:y1sol} affects the
coefficients in all subleading contributions $y_2(z)$, $y_3(z)$ etc.
The intersect of the locus of roots with the negative real axis in the limit
$L\rightarrow \infty$ is now given by
\be
z_{\rm c} = -\frac{1}{\sqrt{a b_{\rm c}}} = - a^{-1/3}
=-\left[\frac{\alpha}{1-\alpha}\right]^{1/3},
\ee
where we have defined
\be
\qquad b_{\rm  c} = \frac{1-\beta_{\rm c}}{\beta_{\rm c}}. 
\ee
The eigenvalue of the transition matrix with the largest nonzero real
part is again determined from \r{eq:EinY}
\bea
{\cal E}_1(L) &=& -\alpha - \frac{1+2z_{\rm c}}{1-z_{\rm c}} -
\frac{4\pi^2}{z_{\rm c}-z_{\rm c}^{-1}}\frac1{L^2} +
\mathcal{O}\left(L^{-3}\right) \nonumber\\ 
&=& -\alpha-\beta_{\rm c} + 
\frac{2}{1+a^{1/3}}  - \frac{4\pi^2}{a^{1/3}-a^{-1/3}} 
\frac{1}{L^2} + \mathcal{O}\left(L^{-3}\right).
\label{eq:energyMII}
\eea
The result \r{eq:energyMII} is valid in the regime
\be
0\leq \alpha <1/2\ ,\qquad \beta_{\rm c} \leq \beta \leq 1,
\ee
and is seen to be independent of $\beta$. Hence in this phase the relaxation to the
stationary state at large times is independent of the extraction rate
at the right-hand boundary of the system.

%%%%%%%%%%%%%%%%%%%%%%%%%%%%%%%%
\section{Maximum Current Phase}
\label{sec:MC}
%%%%%%%%%%%%%%%%%%%%%%%%%%%%%%%%

In the maximum current phase $\alpha,\beta>1/2$ the above
analysis of the integro-differential equation \r{eq:intY} breaks
down. The primary reason for this is that the locus of roots now
closes at $z_{\rm c}=-1$ for $L\to\infty$ and this precludes a Taylor
expansion of the kernel $K(w,z)$ around $w=\xi$. A more detailed
discussion of the complications arising from $z_{\rm c}=-1$ is presented in \ref{app:abel}. 

In order to determine the large-$L$ behaviour of the eigenvalues of
the transition matrix with the largest real parts in the maximum
current phase, we have resorted to a direct numerical solution of the
Bethe ansatz equations \r{eq:pasep_eq} for lattices of up to $L=2600$
sites. In order to facilitate a finite-size scaling analysis it is
necessary to work with quadruple precision (32 digits in C) at large $L$.
We find that the leading behaviour of the spectral gap in the limit
$L\to\infty$ in the maximum current phase is independent of the
boundary rates $\alpha$ and $\beta$.  

%%%%%%%%%%%%%%%%%%%%%%%%%%%%%%
\subsection{Leading behaviour}
%%%%%%%%%%%%%%%%%%%%%%%%%%%%%%

\begin{figure}[ht]
\centerline{
\begin{picture}(150,100)
\put(0,0){\epsfig{width=150pt,file=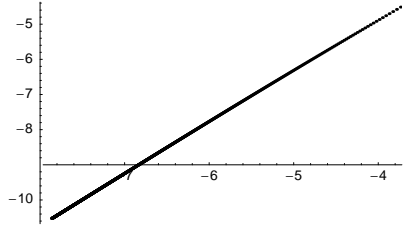}}
\end{picture}}
\caption{Double logarithmic plot of $-{\cal E}_1(L)$ as a function of
$1/L$ for $\alpha=\beta=0.7$}  
\label{fig:logE}
\end{figure}

In Figure~\ref{fig:logE} we plot the numerical results for eigenvalue
${\cal E}_1$ of the first excited state for $\alpha=\beta=0.7$ as a
function of the inverse system size $L^{-1}$ on a
double-logarithmic scale. The almost straight line suggests an
algebraic behaviour at large $L$
\be
{\cal E}_1(L)\sim -eL^{-s}\qquad (L\rightarrow\infty).
\ee
A simple least-square fit of the graph in Figure~\ref{fig:logE} in
the region $2200 \leq L \leq 2600$ to a straight line gives a slope of
$s\approx 1.493$, which is close to the expected value $3/2$. 

\begin{figure}[ht]
\centerline{
\begin{picture}(150,100)
\put(0,0){\epsfig{width=150pt,file=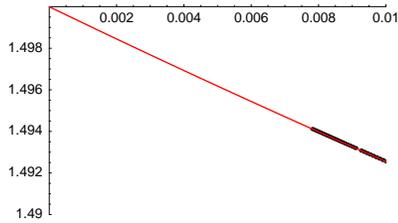}}
\end{picture}}
\caption{Polynomial fit to the exponents $s_k$ of \r{Efit} plotted
against $1/k$ for $\alpha=\beta=0.7$. Extrapolation gives an exponent
$s_\infty = 1.5$.}  
\label{fig:bMCfit}
\end{figure}

A better result is obtained by extrapolating our finite-size data
as follows. We first divide the data set for ${\cal E}_1(L)$ into bins
containing 11 data points each. The $k^{\rm th}$ bin $B_k$ is defined
by taking $20 k\leq L\leq 20(k+1)$. Within each bin we fit the data
for ${\cal E}_1(L)$ to a functional form
\be
{\cal E}_1(L) \approx -e_k L^{-s_k}.
\label{Efit}
\ee
We have implemented this procedure for $100 \leq k \leq 128$ and
obtained a sequence of exponents $s_k$. We observe that the following
least-square fit of the sequence $s_k$ to a polynomial gives excellent
agreement 
\be
s_k \approx 1.4999949 - 0.7822533 k^{-1} + 3.8014387 k^{-2}.
\ee
Finally, we extrapolate to $k=\infty$ and obtain
\be
s_{\infty}\approx 1.4999949.
\ee
This is very close to the expected result $s=3/2$. The polynomial fit
as well as the extrapolation is shown in Figure~\ref{fig:bMCfit}.

%%%%%%%%%%%%%%%%%%%%%%%%%%%%%%%%%%%
\subsection{Subleading corrections}
%%%%%%%%%%%%%%%%%%%%%%%%%%%%%%%%%%%

\begin{figure}[ht]
\centerline{
\begin{picture}(150,100)
\put(0,0){\epsfig{width=150pt,file=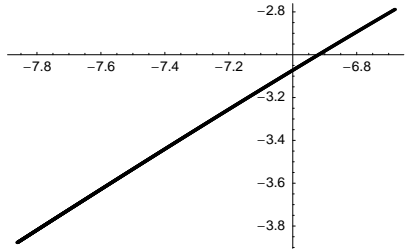}}
\end{picture}}
\caption{Double logarithmic plot of $\Delta_L$ as a function of $1/L$ for 
  $\alpha=\beta=0.7$ and $800\leq L \leq 2600$.} 
\label{fig:logEsub}
\end{figure}

Having established that the leading behaviour of ${\cal E}_1(L)$ at large
$L$ is as $L^{-3/2}$, we now turn to subleading corrections. Assuming
that
\be
{\cal E}_1(L)\sim -e L^{-3/2} - f L^{-d}\ ,\qquad (L\to\infty),
\label{eq:subleading}
\ee
we wish to determine the value of the exponent $d$. To this end, we
define the sequence
\be
\Delta_L = \frac{L}{2} \left[(L+2)^{3/2} {\cal E}_1(L+2) 
- L^{3/2} {\cal E}_1(L)\right).
\ee
If \r{eq:subleading} is correct, we expect $\Delta_L$ to be proportional to
$L^{3/2-d}$ at large $L$. In Figure~\ref{fig:logEsub} we plot
$\Delta_L$ as a function of $L^{-1}$ in a double-logarithmic plot. The
result is well approximated by a straight line with slope $0.95$,
which suggests that the exponent of the subleading corrections is
$d=5/2$. 

\begin{figure}[ht]
\centerline{
\begin{picture}(150,100)
\put(0,0){\epsfig{width=150pt,file=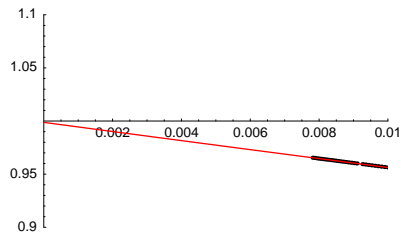}}
\end{picture}}
\caption{Polynomial fit to the exponents $s_k$ in \r{Deltafit} plotted
against $1/k$ for $\alpha=\beta=0.7$.}  
\label{fig:b2MCfit}
\end{figure}

A more accurate estimate of the exponent can be obtained by
extrapolating the finite-size data along the same lines as before.
We group the data for $s_L$ into bins of 11 points each and carry
out a least-square fit for each bin to a functional form
\be
\Delta_L \approx e_k L^{-s_k}.
\label{Deltafit}
\ee
The resulting sequence of exponents $s_k$ is described very
well by the polynomial least-square fit
\be
s_k \approx 0.998755255 - 4.329298118 k^{-1} + 8.363104997 k^{-2}.
\ee
Extrapolation gives $s_\infty\approx 0.998755255$, which is very close to
$1$. This suggests that the subleading corrections to ${\cal E}_1$
indeed scale like $L^{-5/2}$.
The polynomial fit to the sequence of exponents $s_k$ and the
extrapolation to $k=\infty$ are shown in Figure~\ref{fig:b2MCfit}.

%%%%%%%%%%%%%%%%%%%%%%%%%%%%%%%%%%%%%%%%%%%%%%%%%%%%%%%
\subsection{Coefficient of the $L^{-3/2}$ term}
%%%%%%%%%%%%%%%%%%%%%%%%%%%%%%%%%%%%%%%%%%%%%%%%%%%%%%%

\begin{figure}[ht]
\centerline{
\begin{picture}(150,100)
\put(0,0){\epsfig{width=150pt,file=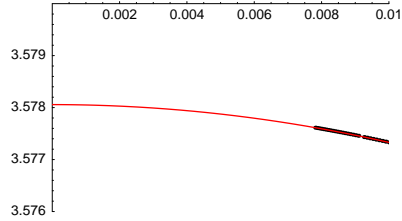}}
\end{picture}}
\caption{Polynomial fit to the coefficients $e_k$ defined in \r{afit} 
plotted against $1/k$ for $\alpha=\beta=0.7$.} 
\label{fig:aMCfit}
\end{figure}

Having established that asymptotically the $L$-dependence of the
eigenvalue is given by
\be
{\cal E}_1(L) \sim -e L^{-3/2} - f L^{-5/2}\qquad (L\rightarrow\infty),
\label{eq:Easymp}
\ee
we now determine the coefficient $e$. We again arrange the data for 
${\cal E}_1(L)$ into bins $20k\leq L\leq 20(k+1)$ and within each bin
perform least-square fits of ${\cal E}_1(L)$ to the functional form
\be
{\cal E}_1(L) \approx -e_k L^{-3/2} - f_k L^{-5/2}\ ,\quad 20k\leq L\leq 20(k+1).
\label{afit}
\ee
As is shown in Figure~\ref{fig:aMCfit}, the resulting sequence of
coefficients $e_k$ for the largest available values of $k$ ($100\leq
k\leq 128$) can be fitted very well to the polynomial
\be
e_k \approx 3.5780576  - 7.2704902 k^{-2}.
\ee

\noindent
Extrapolation to $k\to \infty$ then gives the following result for
the eigenvalue of the first excited state of the TASEP in the MC phase
\be
{\cal E}_1(L) \approx -3.578\, L^{-3/2} + \mathcal{O}(L^{-5/2}).
\label{eq:eoflMC}
\ee
As we have already mentioned, the behaviour \r{eq:eoflMC} is in fact
independent of $\alpha$ and $\beta$ throughout the maximum current
phase and the coefficient is a universal number.

The numerical value for the gap is smaller than that of the 
half-filled TASEP on a ring, where it is found that 
${\cal E}_{1,\rm ring}(L)\sim -6.509\ldots L^{-3/2}$ 
\cite{BAring1,BAring2,GoliM04}. 

%%%%%%%%%%%%%%%%%%%%%%%%%%%%%%%%%%%%%%%%%%%%%%%%%%%%%%%
\subsubsection{Extrapolation Procedure}
%%%%%%%%%%%%%%%%%%%%%%%%%%%%%%%%%%%%%%%%%%%%%%%%%%%%%%%
In order to assess the accuracy of the numerical extrapolation
procedure we have employed above, it is instructive to consider the
analogous analysis on the coexistence line. Here the exact analytical
value for the spectral gap is known. For $\alpha=\beta=0.3$ 
equation \r{eq:energyCL} gives 
\be
{\cal E}_1(L)= -5.1815423\ldots L^{-2} + \mathcal{O}(L^{-3}).
\label{eq:coeffCL}
\ee
We have computed ${\cal E}_1(L)$ on the coexistence line by a direct
numerical solution of the Bethe ansatz equations for lattices of up
to $L=1800$ sites. Fitting ${\cal E}_1(L)$ to the form
\be
{\cal E}_1(L) \approx -e_k L^{-s_k}\ ,\quad 20k\leq L\leq 20(k+1),
\ee
and using the same analysis as above, but for the somewhat smaller values
$61\leq k\leq 88$, we find that the sequence of exponents $s_k$ is well
approximated by the polynomial
\be
s_k \approx 2.0000967 - 0.3772271 k^{-1} + 0.7953103 k^{-2}.
\ee
Extrapolation to $k=\infty$ gives excellent agreement with the exact
result $s=2$.
Similarly, a least-square fit of the sequence of coefficients $e_k$ to
a polynomial in $1/k$ gives
\be
e_k \approx 5.1814722 - 0.0002060 k^{-1} + 0.0000782 k^{-2}.
\ee
The extrapolated value $e_\infty$ agrees with the exact result \r{eq:coeffCL}
to five significant digits. This is quite satisfactory.

%%%%%%%%%%%%%%%%%%%%%%%%%%%%%%%%%%%%%%%%%%%%%
\section{Other Gaps and Complex Eigenvalues}
%%%%%%%%%%%%%%%%%%%%%%%%%%%%%%%%%%%%%%%%%%%%%

Having established the large-$L$ behaviour of the eigenvalue of $M$
corresponding to the lowest excited state, we now turn to higher
excited states. As $M$ is not Hermitian, its eigenvalues are in
general complex. A complex eigenvalue in turn leads to interesting
oscillatory behaviour in the time evolution.
Aspects of such behaviour have been discussed for the KPZ equation
\cite{colaiori,praehofer} 
%and the TASEP on a ring \cite{praehofer}.

%%%%%%%%%%%%%%%%%%%%%%%%%%%%%
\subsection{Massive Phase I: $\alpha< \alpha_{\rm c}$, 
$\beta<\beta_{\rm c}$, $\alpha\neq\beta$}
\begin{figure}[ht]
\centerline{
\begin{picture}(300,100)
\put(0,0){\epsfig{width=300pt,file=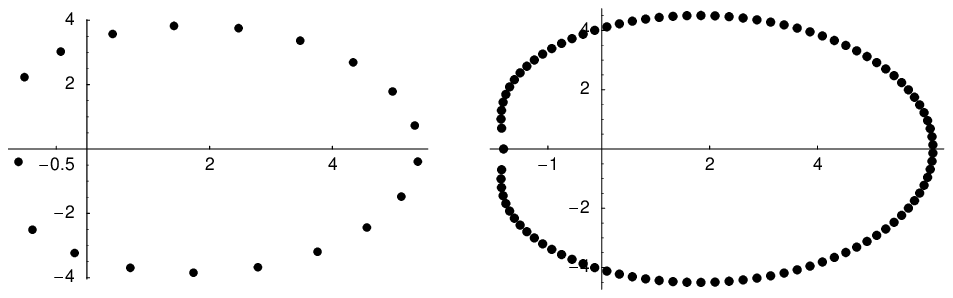}}
\put(0,0){(a)}
\put(150,0){(b)}
\end{picture}}
\caption{Reciprocal root distributions corresponding to the hole state
for $\alpha=0.3$ and $\beta=0.4$; (a) $L=2n=20$\quad (b) $L=2n=100$.} 
\label{fig:2ndexc0304}
\end{figure}

%%%%%%%%%%%%%%%%%%%%%%%%%%%%%
Here the next lowest excitation is obtained by choosing the integers
$I_j$ in the Bethe ansatz equations \r{eq:logtasepBAE} as 
\be
I_j = -L/2 +j \quad\mathrm{for}\quad j=1,\ldots,L-2\qquad I_{L-1} =
L/2.
\ee
This choice of integers $I_j$ corresponds to a ``hole'' between the
last two roots. Hence we will refer to this state as a ``hole state''. 
As this choice is asymmetric with respect to the
interchange $ j \leftrightarrow L-j$, we expect the corresponding
eigenvalue to be complex. This is indeed the case for small system
sizes $L$. However, as the system size increases, the eigenvalue
becomes real at a certain \textit{finite} value of $L$. This comes
about in the following way. For small $L$, the last root has a finite
imaginary part and the distribution of the other roots is asymmetric
with respect to the real axis, see Figure~\ref{fig:2ndexc0304}(a). 
As $L$ increases the last root approaches the negative real axis until
above a critical value of $L$ its imaginary part vanishes. The other roots are
then arranged in complex conjugate pairs, so that the corresponding
eigenvalue becomes real. An example of this is shown in Figure~\ref{fig:2ndexc0304}(b).
Further details regarding this phenomenon are given in \ref{app:finestruc}.

Assuming that the root distribution remains as in Figure~\ref{fig:2ndexc0304}(b) in the
limit $L\rightarrow\infty$, we can compute the corresponding eigenvalue in the following way. As
$z_{L-1}$ is an isolated root, we first write the counting function in
\r{eq:logtasepBAE} as
\be
\i Y_L(z) = g_{\rm}(z) + \frac{1}{L} (g_{\rm b}(z) + K(z_{L-1},z)) 
+ \frac{1}{L} \sum_{l=1}^{L-2} K(z_l,z).
\label{Yexc2}
\ee
We can now follow the same procedure as in Section \ref{se:BAanalysis}
and turn this into an integro-differential equation of the form
\r{eq:intY}. Importantly, the endpoints of the integration contour are
again complex conjugates of one another. Compared to the equation for
the lowest excitation the driving term has an extra contribution of
order ${\cal O}(L^{-1})$
%
%\bea
%g_{\rm b}(z) &=& \ln \left(\frac{z}{1-z^2}\right) +
%\ln\left(z+a\right) + \ln\left(z+b\right) \nonumber\\
%&& + \ln (1/z_{L-1}-z) ,
%\label{gbredef}
%\eea
%
and the boundary conditions that determine the endpoints $\xi$ and
$\xi^*$ of the contour now read
\be
Y_L(\xi^*) = -\pi+\frac{\pi}{L},\qquad Y_L(\xi) = \pi - \frac{3\pi}{L}.
\label{xiredef}
\ee
The root $\zeta=z_{L-1}$ is determined by the requirement that
\be
Y_L(\zeta) = \pi.
\ee

If we expand $\xi$, $\xi^*$ and $Y_L$, but not $\zeta$, in powers of $L^{-1}$, we can use the
results of Section \ref{se:LandH}. With
\be
Y_L(z) = \sum_{n=0}^\infty L^{-n} y_n(z),
\ee
 we find the same solutions for $y_0$, $y_2$ and $y_3$ as in
\r{eq:Z0sol}, \r{eq:y2sol} and \r{eq:y3sol}. However, due to the
additional term proportional to $L^{-1}$ in \r{Yexc2}, the solution
for $y_1$ differs from that in \r{eq:y1sol} and is given by 
\bea
y_1(z) &=& - \i \ln\left[ -\frac{z}{z_{\rm c}} \frac{1-z_{\rm c}^{2}}{1-z^{2}}
\left(\frac{z_{\rm c}-z_{\rm c}^{-1}}{z-z_{\rm c}^{-1}}\right)^{-\i\lambda_1}
\frac{z+a}{z_{\rm c}+a} \frac{z+b}{z_{\rm 
    c}+b}\right] \nonumber\\
&& {}-\i\ln\left[ - \frac{z-1/\zeta}{z_{\rm 
    c}-1/\zeta} \right] +\kappa_1-\i\ln \left(-\zeta^{-1} ab (-z_{\rm c})^{-\i\lambda_1}\right)
\label{eq:y1solexc2}.
\eea
Only now will we expand $\zeta$,
\be
\zeta = \zeta_0 + \sum_{n=1}^\infty L^{-n} \tilde{\delta}_n,
\label{zetaexp}
\ee
and use the definitions of the points $\xi$, $\xi^*$ and $\zeta$ above. Employing again
the expansion \r{eq:expansion} and \ref{ap:Mcoef}, we find that the intersect of
the solution curve with the negative real axis in the limit
$L\rightarrow \infty$ is given, as before, by
\be
z_{\rm c} = -\frac{1}{\sqrt{a b}}
\label{zc2}.
\ee
Furthermore, in leading order, the isolated root $\zeta$ tends to $z_{\rm c}$ in the
limit, i.e. $\zeta_0=z_{\rm c}$. Putting everything together we
finally find that the eigenvalue for this solution is given by  
\be
{\cal E}_2(L) =-\alpha-\beta - \frac{2z_{\rm c}}{1-z_{\rm c}} -
\frac{4\pi^2}{z_{\rm c}-z_{\rm c}^{-1}} \frac1{L^2} +\mathcal{O}(L^{-3}),
\label{eq:energyMI2}
\ee
with $z_{\rm c}$ given in \r{zc2}. It is instructive to compare ${\cal
  E}_2(L)$ to the gap of the first excited state ${\cal E}_1(L)$. The
latter exhibits a non-analytic change at
  $\beta=\beta_{\rm c}$. Interestingly, we find that to order
  $\mathcal{O}(L^{-2})$ 
\be
\lim_{\beta\uparrow\beta_{\rm c}}{\cal E}_2(L)=\lim_{\beta\downarrow\beta_{\rm c}}{\cal
  E}_1(L) \neq \lim_{\beta\uparrow\beta_{\rm c}}{\cal E}_1(L)\ .
\ee
In conclusion we find that in the massive phase $M_{\rm I}$ also the second
gap is real and given by \r{eq:energyMI2}. The second gap may be a
complex conjugate pair for small values of the system size. However, this pair
merges at a \emph{finite} value of $L$ producing two real
eigenvalues, the lowest of which is given by \r{eq:energyMI2}. This observation is consistent 
with the results of Dudzinski and Sch\"utz \cite{DudzS00}, who computed the spectrum
on the coexistence line $\alpha=\beta$ for small system sizes.

%%%%%%%%%%%%%%%%%%%%%%%%%%%%%
\subsection{Coexistence line}
%%%%%%%%%%%%%%%%%%%%%%%%%%%%%

As we have seen above, the second gap of the TASEP is real 
for large $L$, even though it may be complex for small system
sizes. On the coexistence line, the leading term of \r{eq:energyMI2}
vanishes and we are left with a second diffusive mode with an
eigenvalue that vanishes as $\mathcal{O}(L^{-2})$. A priori we expect
a whole band of diffusive modes on the coexistence line, some of which
should be given by the domain wall theory of \cite{DudzS00},
\be 
{\cal E}_n(L) = -\frac{n^2\pi^2}{z_{\rm c} - z_{\rm c}^{-1}}
\frac{1}{L^2} + \mathcal{O}(L^{-3}),
\ee
with $z_{\rm c}$ given by \r{zc2}. It is clearly
of interest to know whether or not there are complex modes as well, which
would result in oscillatory behaviour in the relaxational dynamics at
large times. 

\begin{figure}[ht]
\centerline{
\begin{picture}(150,100)
\put(0,0){\epsfig{width=150pt,file=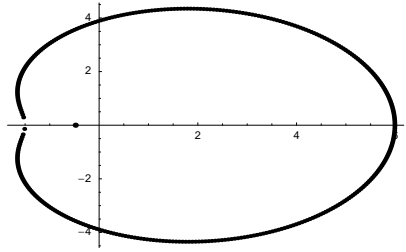}}
\end{picture}}
\caption{Reciprocal root distributions corresponding to the first complex
  eigenvalue ${\cal E}_{\rm c}(L)$ for $\alpha=\beta=0.3$ and $L=1860$.}
\label{fig:3rdexc0303}
\end{figure}

We observed numerically, that the first complex excitation that does
not become real for a finite value of $L$ in the massive phase $M_{\rm
  I}$ corresponds to the choice of integers 
\bea
I_j &=& -L/2 +j \quad\mathrm{for}\quad j=1,\ldots,L-3 \nonumber\\
I_{L-2} &=& L/2-1,\quad I_{L-1} = L/2,
\label{Ichoice3}
\eea
i.e. there is one hole between the second and third last roots.
The corresponding (reciprocal) root distribution again has an isolated
root on the negative real axis, see Figure~\ref{fig:3rdexc0303}, which
now lies inside the contour of integration. 

Proceeding as in the preceding section, we write the counting function as
\be
\i Y_L(z) = g_{\rm}(z) + \frac{1}{L} (g_{\rm b}(z) + K(z_{L-1},z) +
\frac{1}{L} \sum_{l=1}^{L-2} K(z_l,z).
\label{Yexc3}
\ee
Turning the sum into an integral we arrive at the following
integro-differential equation
\bea
\i\,Y_L(z) &=& g(z) + \frac{1}{L} \left(g_{\rm b}(z) + K(\zeta_1,z) -
K(\zeta_2,z)\right)  \nonumber \\  
&& \hspace{-1cm} +\frac{1}{2\pi}
\int_{\xi_2}^{\xi_1} K(w,z) Y'_L(w) \d w \nonumber\\
&& \hspace{-1cm}+ \frac{1}{2\pi} \int_{C_1} \frac{K(w,z)Y'_L(w)}{1-\e^{-\i L
Y_L(w)}}\, \d w + \frac{1}{2\pi} \int_{C_2} \frac{K(w,z)Y'_L(w)}{
\e^{\i L Y_L(w)}-1}\,\d w.
\label{eq:intY3}
\eea
Here, $\zeta_1 =z_{L-1}$ is the isolated root and $\zeta_2$
corresponds to the position of the hole. The contribution due to the
latter needs to be
subtracted in order to cancel the contribution arising from the
integral. The values of $\zeta_1$, $\zeta_2$  as well as the endpoints
$\xi_1$ and $\xi_2$ of the curve are determined self-consistently by
the ``boundary conditions'' 
\be
\renewcommand{\arraystretch}{2.2}
\begin{array}{ll}
\displaystyle Y_L(\xi_2) = -\pi+\frac{\pi}{L}, \qquad & \displaystyle Y_L(\xi_1) = \pi -
\frac{\pi}{L}, \\
\displaystyle Y_L(\zeta_1) = \pi, & \displaystyle Y_L(\zeta_2) = \pi - \frac{4\pi}{L}.
\end{array}
\label{xiredef3}
\ee

Assuming an expansion of the form 
\be
Y_L(z) = \sum_{n=0}^\infty L^{-n} y_n(z),
\label{Yexp}
\ee
we can utilize the results of Section \ref{se:LandH}. If we
furthermore assume that $\zeta_1^{-1}$ lies inside the contour of
integration, we find that $y_0(z)$ is again given by \r{eq:Z0sol} and
obtain the following result for $y_1(z)$
\bea
y_1(z) &=& - \i \ln\left[ -\frac{z}{z_{\rm c}} \frac{1-z_{\rm c}^{2}}{1-z^{2}}
\left(\frac{z_{\rm c}-z_{\rm c}^{-1}}{z-z_{\rm c}^{-1}}\right)^{-\i\lambda_1}
\frac{z+a}{z_{\rm c}+a} \frac{z+b}{z_{\rm 
    c}+b}\right] \nonumber\\
&& {}-\i\ln\left[ \frac{z-1/\zeta_1}{z_{\rm 
    c}-1/\zeta_1} \frac{z-\zeta_1}{z_{\rm 
    c}-\zeta_1} \right] +\i\ln\left[ - \frac{z-1/\zeta_2}{z_{\rm 
    c}-1/\zeta_2} \right] \nonumber\\
&& {} +\kappa_1-\i\ln \left(-\zeta_2 ab (-z_{\rm c})^{-\i\lambda_1}\right)
\label{eq:y1solexc3}.
\eea
It is important to note, that expression \r{eq:y1solexc3} has been
derived under the assumption that $z^{-1}$ lies outside the contour of
integration\footnote{Otherwise the term $\ln(1-wz)$ in the kernel $K(w,z)$
would produce an additional contribution in the integro-differential
equation for $y_1(z)$.}. Hence, we cannot use \r{eq:y1solexc3} to
determine $\zeta_1$ via \r{xiredef3}. However, it turns out that
\r{eq:y1solexc3} already contains sufficient information for
determining the leading order of ${\cal E}_{\rm c}$.

Using the definitions of the points $\xi_1$, $\xi_2$ and $\zeta_2$
above, the expansion \r{eq:expansion} and \ref{ap:Mcoef}, we find that
the intersect of the solution curve with the negative real axis in the
limit $L\rightarrow \infty$ is now given by
\be
z_{\rm c} = -(a b)^{-1/4}.
\label{zc3}
\ee
The ${\cal O}(1)$ term of the eigenvalue corresponding to the choice
\r{Ichoice3} is real
\be
{\cal E}_{\rm c}(L)= -\alpha-\beta - \frac{1+3z_{\rm c}}{1-z_{\rm c}} + o(1),
\label{eq:energyMI3}
\ee
where $z_{\rm c}$ given in \r{zc3}. The subleading corrections are 
complex. However, the ${\cal O}(1)$ contribution \r{eq:energyMI3} 
to the eigenvalue does \emph{not} vanish on the coexistence line.
Hence this excitation does not play an important role at large times.
We have checked that for $\alpha=\beta=0.3$ both \r{zc3} and 
\r{eq:energyMI3} agree well with a direct numerical solution of the
Bethe ansatz equations. 
Based on the above analysis, we conjecture that for all gapless
excitations on the coexistence line with eigenvalues that scale with
system size as
\be
{\cal E}=-eL^{-2}+{\cal O}(L^{-3}),
\ee
the coefficients $e$ are real. In other words, the dominant large-time
relaxation on the coexistence line does not have oscillatory modes.

%%%%%%%%%%%%%%%%%%%%%%%%%%%%%%%%%%
\subsection{Maximum Current phase}
%%%%%%%%%%%%%%%%%%%%%%%%%%%%%%%%%%

Comparing again the numerical solutions of the Bethe ansatz equations to a
direct numerical determination of the eigenvalues of the transition
matrix for small system sizes $L$ we find that the lowest
excitations with non-vanishing imaginary part are characterized by the
set of integers
\be
I_j = -L/2 +j \quad\mathrm{for}\quad j=1,\ldots,L-2\qquad I_{L-1} =
L/2,
\label{complexrootdistr}
\ee
or
\be
I_j = -L/2 +j \quad\mathrm{for}\quad j=2,\ldots,L-1\qquad I_{1} = -L/2.
\ee
For the first choice the imaginary part of the eigenvalue is positive,
whereas the second yields the complex conjugate eigenvalue. In
contrast to the corresponding excitations in the massive phase $M_{\rm
  I}$, we find that the gaps in the MC phase remain complex for all
values of $L$.

We denote the complex eigenvalue corresponding to
\r{complexrootdistr} by ${\cal E}_{\rm c}(L)$. We have determined
${\cal E}_{\rm c}(L)$ from a numerical solution of the Bethe ansatz
equations for lattices of up to $L=1200$ sites and find its leading
large $L$ behaviour to be independent of the rates $\alpha$ and
$\beta$. In what follows we therefore fix $\alpha=\beta=0.7$, in the
understanding that the results for the large-$L$ asymptotics of the
eigenvalue of $M$ are universal throughout the maximum current phase. 

\begin{figure}[ht]
\centerline{
\begin{picture}(150,100)
\put(0,0){\epsfig{width=150pt,file=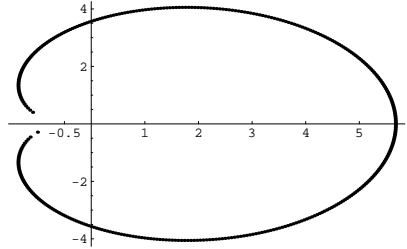}}
\end{picture}}
\caption{Reciprocal root distributions corresponding to the lowest
  complex eigenvalue ${\cal E}_{\rm c}(L)$ for $\alpha=\beta=0.7$ and $L=2n=400$.} 
\label{fig:complexMCroots}
\end{figure}

The distribution of reciprocal Bethe roots  for $L=400$ and
$\alpha=\beta=0.7$ is shown in Figure~\ref{fig:complexMCroots}.  
We observe that the root distribution is quite similar to that of the
second lowest eigenvalue in Figure~\ref{fig:roots}. The main
difference is that now there is a slight gap between the last two
roots. 

\begin{figure}[ht]
\centerline{
\begin{picture}(350,100)
\put(0,0){\epsfig{width=320pt,file=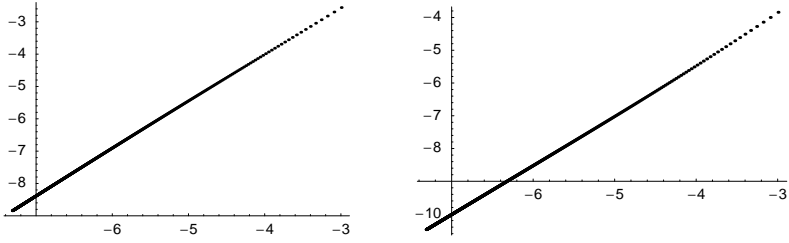}}
\end{picture}}
\caption{Double logarithmic plots of the real and imaginary 
parts ${\cal E}_{\rm c}$ as functions of $1/L$.} 
\label{fig:complex_exc}
\end{figure}

Based on the similarity of the root distribution to the one of the
lowest excited state, we expect that both real and imaginary parts
of ${\cal E}_{\rm c}(L)$ will be proportional to  $L^{-3/2}$ for large
systems. This is indeed the case, as is shown in Figure~\ref{fig:complex_exc}, where ${\rm
  Re}({\cal E}_{\rm c}(L))$ and ${\rm 
  Im}({\cal E}_{\rm c}(L))$ are plotted as functions of $L^{-1}$ on a double
logarithmic scale for $\alpha=\beta=0.7$. Both curves are very close
to straight lines with slope $3/2$. 

\begin{figure}[ht]
\centerline{
\begin{picture}(300,100)
\put(0,0){\epsfig{width=300pt,file=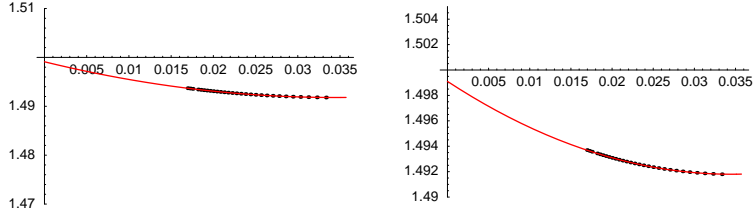}}
\end{picture}}
\caption{Fits to the exponents $s_k$ and $t_k$ in \r{complexEfit} of
the real and imaginary parts of ${\cal E}_{\rm c}(L)$ plotted against $1/k$
for $\alpha=\beta=0.7$.} 
\label{fig:complexbMCfit}
\end{figure}

In order to determine the asymptotic form of ${\cal E}_{\rm c}(L)$ more
accurately, we repeat the analysis we employed for the lowest excited
state in the maximum current phase in section \ref{sec:MC}. We group
the data points for ${\cal E}_{\rm c}(L)$ in bins containing 11 points each
and within each bin perform least square fits to
\bea
{\rm Re}({\cal E}_{\rm c}(L)) \approx {\rm Re}(e_k) L^{-s_k}\ ,\nn
{\rm Im}({\cal E}_{\rm c}(L)) \approx {\rm Im}(e_k) L^{-t_k}\ , \quad 20k\leq
L\leq 20(k+1). 
\label{complexEfit}
\eea
The resulting sequences of exponents $s_k$ and $t_k$ in the range
$30\leq k \leq 59$ are well approximated by the polynomials
\bea
s_k &\approx& 1.4997697 - 0.737888549 k^{-1} + 3.369432984 k^{-2}\ ,\nn
t_k &\approx& 1.4991041 - 0.423377041 k^{-1} + 6.133640280 k^{-2}.
\eea
The extrapolated values at $k=\infty$ are very close to $3/2$.
The fits to the data as well as the extrapolation to $k=\infty$ are
show in Figure~\ref{fig:complexbMCfit}. 

\begin{figure}[ht]
\centerline{
\begin{picture}(300,100)
\put(0,0){\epsfig{width=300pt,file=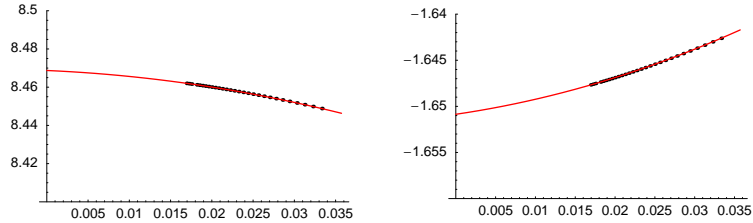}}
\end{picture}}
\caption{Polynomial fits to the real and imaginary parts of the
 coefficient $e_k$ in \r{complexafit} plotted against $1/k$ for 
$\alpha=\beta=0.7$.} 
\label{fig:complexaMCfit}
\end{figure}

In order to determine the coefficients of the $\mathcal{O}(L^{-3/2})$
terms, we carry out least square fits of the binned data for ${\cal
  E}_{\rm c}(L)$ to
\be
{\cal E}_{\rm c}(L) \approx -e_k L^{-3/2} -f_k L^{-5/2}\ ,\quad
20k\leq L\leq 20(k+1).
\label{complexafit}
\ee
The resulting sequence of coefficients $e_k$ is well described by the
polynomial fits (see Figure~\ref{fig:complexaMCfit})
\be
\begin{array}{ll}
{\rm Re}(e_k) \approx 8.4687424 - 0.1886671 k^{-1} -12.277915 k^{-2},\\
{\rm Im}(e_k) \approx -1.6508588 + 0.1263865 k^{-1} + 3.6333854 k^{-2}.
\end{array}
\ee
\noindent
Extrapolation to the limit $k\to \infty$ then gives the following result for
the energy of the first complex excited state of the TASEP in the MC phase
\be
{\cal E}_{\rm c}(L) \approx -(8.47 - 1.65\,\i)  \, L^{-3/2} +
\mathcal{O}(L^{-5/2}). 
\label{eq:complexE}
\ee
The result \r{eq:complexE} can be compared to excited states with
complex eigenvalues in the half-filled 
\footnote{It is natural to compare the the half-filled case, as the
average bulk density of the TASEP with open boundaries in the maximum
current phase is $1/2$.} 
TASEP on a ring. We are not aware of any explicit results in the
literature on complex eigenvalues, but they can be easily determined
by employing the method of \cite{GoliM04}. We summarize some results
in \r{app:ring}. The lowest excited state with complex eigenvalue
found to be
\be
{\cal E}_{{\rm c},{\rm ring}}(L) \sim -(17.1884\ldots - 5.43662\ldots\,\i)  \,
L^{-3/2}. 
\ee
%%%%%%%%%%%%%%%%%%%%%%%%%%%%%%%%%%%%%%
\section{Summary and Conclusions}
\label{se:summ}
%%%%%%%%%%%%%%%%%%%%%%%%%%%%%%%%%%%%%%

In this work we have used Bethe's ansatz to diagonalize the transition
matrix $M$ for arbitrary values of the rates $p$, $q$, $\alpha$,
$\beta$, $\gamma$ and $\delta$ that characterize the most general
PASEP with open boundaries. The resulting Bethe ansatz equations
\r{eq:pasep_en}, \r{eq:pasep_eq} describe the \emph{complete}
excitation spectrum of $M$.

We have carried out detailed analyses of the Bethe ansatz equations
for the limiting cases of symmetric and totally asymmetric exclusion
and determined the exact asymptotic behaviour of the spectral gap for
large lattice lengths $L$. This gap determines the long time ($t\gg 
L$) dynamical behaviour of the TASEP. We emphasize that care has to be 
taken regarding time scales, and that our results below are not valid
at intermediate times where the system behaves as
for periodic boundary conditions \cite{Schuetz00}.

%%%%%%%%%%%%%%%%%%%%%%%%%%%%%%%%%%%%%%
\subsection{Dynamical phase diagram}
\begin{figure}[ht]
\begin{center}
\epsfxsize=0.6\textwidth
\epsfbox{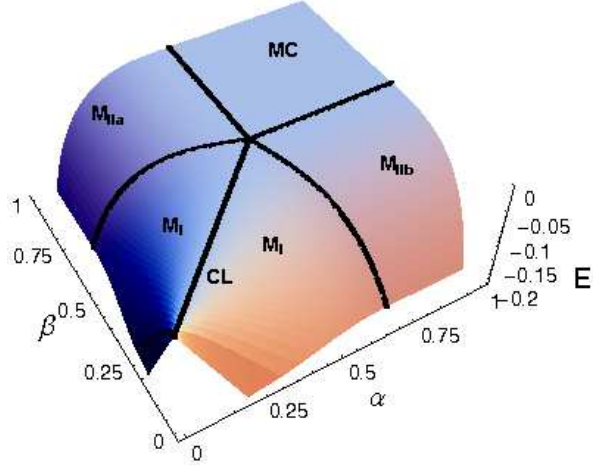}
\caption{Dynamic phase diagram determined by the first eigenvalue gap
  of the TASEP. M$_{\rm I}$, M$_{\rm IIa}$ and M$_{\rm IIb}$ are
  massive phases, CL denotes the critical coexistence line and MC the
  critical maximal current phase.}
\label{fig:phase}
\end{center}
\end{figure}

%%%%%%%%%%%%%%%%%%%%%%%%%%%%%%%%%%%%%%
In the case of totally asymmetric exclusion and $\gamma=\delta=0$, we
found that there are three regions in parameter space where the
spectral gap is finite and the stationary state is approached
exponentially fast. In addition there is one region (maximum current
phase) and a line (coexistence line) where the gap vanishes as
$L\rightarrow\infty$. The resulting dynamical phase diagram 
\footnote{We use the term ``phase'' to indicate a
region in parameter space characterised by a particular type of
relaxational behaviour.} is shown in Figure~\ref{fig:phase}.

Recalling that 
\be
\beta_{\rm c}=\left[1+\left(\frac{\alpha}{1-\alpha}\right)^{1/3}\right]^{-1},\qquad
\alpha_{\rm c}=\left[1+\left(\frac{\beta}{1-\beta}\right)^{1/3}\right]^{-1},
\ee
the leading asymptotic values of the spectral gap in the various regions of the phase diagram
of the TASEP are as follows:
\bigskip

\noindent{\underline{\it Massive Phase M$_{\rm I}$:
$\alpha< \alpha_{\rm c}$, $\beta<\beta_{\rm c}$, $\alpha\neq\beta$}}
\be
{\cal E}_1(L)= -\alpha-\beta + 
\frac{2}{1+\sqrt{(1-\alpha)(1-\beta)/\alpha\beta}} + \mathcal{O}(L^{-2}).
\label{eq:E_MI}
\ee
The spectral gap does not vanish as $L\rightarrow\infty$, which
implies a finite correlation length and an exponentially fast approach
to stationarity.\hfill\vskip .2cm

\noindent{\underline{\it High-Density Phase M$_{\rm IIb}$: 
$\beta< 1/2$, $\alpha_{\rm c}<\alpha$}}
\be
{\cal E}_1(L) = -\alpha_{\rm c}-\beta + \frac{2}{1+
\left[(1-\beta)/\beta\right]^{1/3}} +
\mathcal{O}(L^{-2}).
\label{eq:E_MII}
\ee
The spectral gap is finite and independent of $\beta$. The
relaxational behaviour is again exponentially fast. 

The gap in the low-density phase M$_{\rm IIa}$:
$\alpha<1/2$, $\beta_{\rm  c}<\beta$ is obtained by the exchange
$\alpha\leftrightarrow\beta$. 
We note that the subdivision of the massive high and low density
phases into $M_{\rm I}$ and $M_{\rm IIa,b}$ is different from the one
suggested on the basis of stationary state properties in
\cite{gunter}. \hfill\vskip .2cm

\noindent\underline{\it Coexistence Line (CL): $\beta=\alpha<1/2$.} 
\be
{\cal E}_1(L) = -\frac{\pi^2 \alpha(1-\alpha)}{1-2\alpha} L^{-2} +
\mathcal{O}(L^{-3}).
\label{eq:E_EW}
\ee
The gap vanishes like $L^{-2}$ for large systems, which corresponds to
a dynamic exponent $z=2$ and indicates relaxational behaviour of a
diffusive type.\hfill\vskip .2cm

\noindent\underline{\it Maximal Current Phase (MC): $\alpha,\beta>1/2$.}
\be
{\cal E}_1(L)\approx -3.578 L^{-3/2} + \mathcal{O}(L^{-5/2}).
\label{eq:e1_open}
\ee
Here the gap is independent of the rates $\alpha$ and $\beta$. The
dynamic exponent $z=3/2$ is indicative of KPZ like behaviour
\cite{KPZ}. We find that the magnitude of ${\cal E}_1(L)$ is
smaller than half that of the lowest excited state for the TASEP with
periodic boundary conditions
\be
{\cal E}_{1,\rm ring}(L)\sim -6.50919\ldots L^{-3/2}.
\ee
The two gaps do not seem to be related in any obvious way. 

It is known \cite{NeergN95} that by varying the bulk hopping
rates one can induce a crossover between a diffusive Edwards-Wilkinson
(EW) scaling regime \cite{EW} with dynamic exponent $z=2$ and a KPZ regime
\cite{KPZ} with $z=3/2$. Here we have shown using exact methods that a
crossover between phases with $z=2$ and $z=3/2$ occurs in the case
where the bulk transition rates are kept constant, but the boundary
injection/extraction rates are varied.

%%%%%%%%%%%%%%%%%%%%%%%%%%%%%%%%%%%%%%%%%%%%%%%
\subsection{Domain wall theory}
%%%%%%%%%%%%%%%%%%%%%%%%%%%%%%%%%%%%%%%%%%%%%%%
As shown in \cite{KSKS,DudzS00} the
diffusive relaxation ($z=2$) is of a different nature than in the EW
regime and is in fact due to the unbiased random walk behaviour of a
shock (domain wall between a low and high density region) with right
and left hopping rates given by
\be
D^{\pm} = \frac{J^{\pm}}{\rho^+ - \rho^-}.
\label{eq:DWTrates}
\ee
Here, $\rho^-=\alpha$ and $\rho^+=1-\beta$ are the stationary bulk
densities in the low- and high-density phases respectively, and the
corresponding currents are given by $J^\pm=\rho^\pm(1-\rho^\pm)$. 
Our results \r{eq:E_MI} and \r{eq:E_EW} for the massive phase $M_{\rm
  I}$  and the coexistence line agree with the relaxation time
calculated in the framework of a domain wall theory (DWT) model in
\cite{DudzS00}. This is in contrast to the massive phases $M_{\rm
  II}$, where the exact result \r{eq:E_MII} differs from the DWT
prediction of \cite{DudzS00}.  More precisely, the DWT predicts that
the relaxational mechanism in the stationary high-density phase
($\beta<1/2$, $\alpha>\beta$) for $\alpha<1/2$ is due to the random
walk of a ``$(0|1)$'' domain wall between a low and a high density
segment. On the other hand, in the high-density phase for
$\alpha>1/2$, the DWT predicts a relaxational mechanism based on
so-called ``$(m|1)$'' domain  walls between a maximum current and a
low-density region. Our results for the gap exhibit a change of
behaviour at $\alpha=\alpha_{\rm c}$ rather than at $\alpha=1/2$. We
therefore propose the following modified DWT. 

\begin{figure}[ht]
\begin{center}
\epsfxsize=0.6\textwidth
\epsfbox{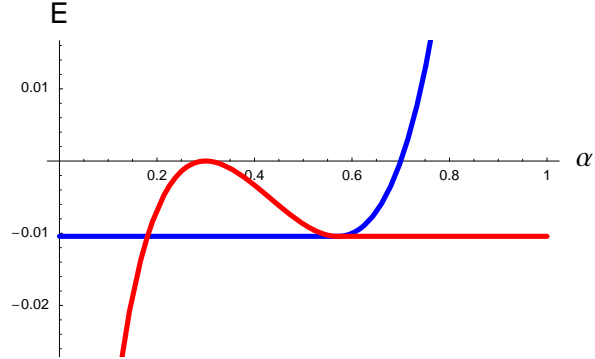}
\caption{The gap as a function of $\alpha$ for $\beta=0.3$. The curve
  corresponds to the function in \r{eq:E_MI} and the horizontal line to
  \r{eq:E_MII}. The gap of the lowest excitation (red online) is a
  combination of the curve ($0< \alpha<\alpha_{\rm c}$) and the line
  ($\alpha_{\rm c}< \alpha<   1$).
}
\label{fig:Ecut}
\end{center}
\end{figure}

We assume that in the high density phase for large $\alpha$ we
can still use the DWT rates as given in \r{eq:DWTrates}, but with an
effective density $\rho^-_{\rm eff}$. In contrast to \cite{DudzS00} we
do not take the effective density equal to that of the maximum current
phase, $\rho^-_{\rm eff}=1/2$, but instead determine it below from
using a monotonicity argument. Consider first Figure~\ref{fig:Ecut}, in which we
plot the first gap as a function of $\rho^-=\alpha$ for constant
$\beta=0.3$. 
\begin{itemize}
\item $\alpha<\beta$\\ 
We are in the low density phase. The gap in the infinite volume is
finite in this region, but upon increasing $\alpha$ we are driven
towards the coexistence line where the gap vanishes and the relaxation
is diffusive. Hence, $\partial \mathcal{E}/\partial\alpha >0$ in this region.  

\item $\alpha > \beta$\\
We are now in the high density region and we expect the gap to be
finite again, decreasing with increasing $\alpha$. Hence, in this
region we should have $\partial \mathcal{E}/\partial\alpha \leq 0$. Initially, this is
indeed the observed behaviour of the graph of \r{eq:E_MI}, but we see
that at $\alpha=\alpha_{\rm c}$ the graph of \r{eq:E_MI} has local
minimum, where the expected behaviour breaks down.
\end{itemize}

Hence, for small values of $\rho^-$ and large values of $\rho^+$ the
domain wall theory of \cite{DudzS00} correctly predicts the gap to be
\be
\mathcal{E}_1(\rho^-,\rho^+) = -D^+-D^- +2\sqrt{D^+D^-}\,,
\ee
which is equal to \r{eq:E_MI}. In the high-density region, there is a critical value $\rho^-_{\rm
  eff}=\alpha_{\rm c}$ beyond which the gap is given by $\mathcal{E}_1(\rho^-_{\rm
  eff},\rho^+)$, where $\rho^-_{\rm eff}$ is determined by
\be
\left. \frac{\partial \mathcal{E}_1(\rho^-,\rho^+)}{\partial \rho^-}
\right|_{\rho^{-}=\rho^-_{\rm eff}} =0.
\ee
The considerations above are of a sufficiently general nature to
remain valid for the wider class of driven diffusive systems that can
be described by effective domain wall theories. We emphasize that the 
DWT as discussed above pertains to the limit $L\rightarrow\infty$
only, and that a more careful analysis is necessary to correctly
produce the difference in finite size behaviour for the phases $M_{\rm
  I}$ and $M_{\rm II}$, see \r{eq:energyMI} and \r{eq:energyMII}. In
particular, the boundary rates generally need to be chosen differently
from the bulk rates. A more detailed comparison of the Bethe ansatz
solution and DWT is possible for the case of the PASEP with $Q\neq 0$,
where DWT becomes exact for certain values of the rates
\cite{KFS}. This is beyond the scope of the present work and will be
discussed in a separate publication \cite{deGE}.

%%%%%%%%%%%%%%%%%%%%%%%%%%%%%%%%%%%%%%%%%%%%
\subsection{Higher excitations}
%%%%%%%%%%%%%%%%%%%%%%%%%%%%%%%%%%%%%%%%%%%%

\noindent{\underline{\it Massive Phase M$_{\rm I}$ and Coexistence
    Line (CL)}}:
\medskip\\
The second gap in this phase is given by
\be
{\cal E}_2(L) =-\alpha-\beta - \frac{2z_{\rm c}}{1-z_{\rm c}} -
\frac{4\pi^2}{z_{\rm c}-z_{\rm c}^{-1}} \frac1{L^2} +\mathcal{O}(L^{-3}),
\ee
where $z_{\rm c}=-1/\sqrt{ab}$. This result is in agreement with the
second gap predicted by the DWT of \cite{DudzS00}. In
particular, on the coexistence line it corresponds to $n=2$ of the
DWT band of diffusive modes on the coexistence line, 
\be 
{\cal E}_n(L) = -\frac{n^2\pi^2}{z_{\rm c} - z_{\rm c}^{-1}}
\frac{1}{L^2} + \mathcal{O}(L^{-3}).
\ee
The first gap given in \r{eq:E_EW} corresponds to $n=1$.

We studied the possibility of an oscillating diffusive mode on the
coexistence line and found that the first excitation with nonzero
imaginary part has a finite gap for $L\to\infty$
\be
{\cal E}_{\rm c}(L)= -\alpha-\beta - \frac{1+3z_{\rm c}}{1-z_{\rm c}} + o(1),
\ee
where $z_{\rm c} = -(a b)^{-1/4}$. Although the subleading terms have
nonzero imaginary parts, the leading term is real and does \emph{not}
vanish on the coexistence line. Hence this excitation does not play an
important role at large times. We conjecture that the leading ${\cal
  O}(L^{-2})$ terms of the eigenvalues of all diffusive modes on the
coexistence line are real. 
\bigskip

\noindent{\underline{\it Maximum Current Phase MC}}
\medskip\\
We determined the energy of the first ``complex'' excited state
in the MC phase by means of a direct numerical solution of the Bethe
ansatz equations. We find it to be independent of $\alpha$ and $\beta$
and given by
\be
{\cal E}_{\rm c}(L) \approx -(8.47 - 1.65\,\i)  \, L^{-3/2} +
\mathcal{O}(L^{-5/2}).
\label{eq:comexc_open}
\ee
Complex eigenvalues give rise to interesting oscillations in
correlation functions. Such behaviour has been observed for the KPZ
equation \cite{colaiori,praehofer}, which is related to the TASEP
with periodic boundary conditions. In \ref{app:ring} we have computed
the low lying complex excitations of the TASEP on the ring using the
method of \cite{GoliM04}. The lowest such excitation is given by 
\be
{\cal E}_{\rm c, ring}(L) \sim -(17.1884\ldots-5.43662\ldots\,\i)L^{-3/2}.
\label{eq:comexc_ring}
\ee
The result \r{eq:comexc_ring} confirms the prediction from the KPZ
equation that both real and imaginary part scale with $L^{-3/2}$
\cite{praehofer}. According to \r{eq:comexc_open} this scaling still holds
when boundaries are present. However, as was the case with the first gap,
although \r{eq:comexc_open} does not depend on the boundary rates
$\alpha$ and $\beta$, its value does not seem to be related to
\r{eq:comexc_ring} in a simple way.
\bigskip

A number of interesting open problems remain. In a forthcoming
publication  \cite{deGE} we determine the spectral gaps for the case
of partially asymmetric diffusion and $\alpha,\beta,\gamma,\delta\neq
0$ from the Bethe ansatz equations \r{eq:pasep_eq}. As we have
discussed, the constraint \r{BAconstr} precludes the determination of
current fluctuations from the Bethe ansatz. It therefore would be
highly desirable to obtain Bethe ansatz equations for the Heisenberg
XXZ chain with the most general open boundary conditions. Furthermore,
the existing Bethe ansatz solution provides information only about the
spectrum, but not the eigenstates of the transition matrix. In order
to study correlations functions \cite{vladb} the knowledge of matrix
elements of the spin operators between left and right eigenstates is
required. In light of this it would be very interesting to construct
the eigenstates from the Bethe ansatz.

\ack
We are grateful to B. Derrida, G.M. Sch\"utz and R. Stinchcombe for
very helpful discussions. This work was supported by the ARC (JdG) and
the EPSRC under grant GR/R83712/01 (FE).

\appendix 

%%%%%%%%%%%%%%%%%%%%%%%%%%%%%%%%%%%%%%%%%%%%%%%%%%%%%%%%%%%%%%
\section{Analysis of the Abel-Plana Formula}
\label{app:abel}
%%%%%%%%%%%%%%%%%%%%%%%%%%%%%%%%%%%%%%%%%%%%%%%%%%%%%%%%%%%%%%
In this appendix we sketch how to extract the finite-size correction
terms from the integral expression
\bea
\i\,Y_L(z) &=& g(z) + \frac{1}{L} g_{\rm b}(z) +\frac{1}{2\pi}
\int_{\xi^*}^{\xi} K(w,z) Y'_L(w)\ \d w \nonumber \\ 
&& \hspace{-1cm}+ \frac{1}{2\pi} \int_{C_1} \frac{K(w,z)Y'_L(w)}{1-\e^{-\i L
Y_L(w)}}\ \d w + \frac{1}{2\pi} \int_{C_2} \frac{K(w,z)Y'_L(w)}{
\e^{\i L Y_L(w)}-1}\ \d w.
\label{eq:intYapp}
\eea
The main contributions to the correction terms in \r{eq:intYapp} comes
from the vicinities of the endpoints $\xi$, $\xi^*$. Along the contour
$C_1$ the imaginary part of the counting function is positive, ${\rm
  Im}(Y_L(w))>0$, whereas along the contour $C_2$ it is negative,
${\rm Im}(Y_L(w))<0$. As a result the integrands decay exponentially
with respect to the distance from the endpoints. In the vicinity of
$\xi$, we therefore expand 
\be
Y_L(w)=Y_L(\xi)+Y_L'(\xi) (w-\xi)+ \ldots\ .
\ee
Assuming that $Y_L'(\xi)$ is ${\cal O}(1)$ (an assumption that will be
checked self-consistently), we find that the leading contribution for
large $L$ is given by
\bea
\frac{1}{2\pi}
\int_{C_1} \frac{K(w,z)Y'_L(w)}{1-\e^{-\i L Y_L(w)}}\ \d w &\sim&
\frac{Y_L'(\xi)}{2\pi} \int_{\xi}^0 \frac{K(w,z)}{1+\e^{-\i L
    Y_L'(\xi)(w-\xi)}}\ \d w\ ,\nn
&& \mbox{} - (\xi \rightarrow \xi^*).
\eea
Here we have used the boundary conditions \r{eq:xidef}. Carrying out
the analogous analysis for the integral along the contour $C_2$, we
arrive at the following expression for the leading contribution of the
last two terms in \r{eq:intYapp} 
\bea
{\cal A} &=& \frac{1}{2\pi} \int_{C_1} \frac{K(w,z)Y'_L(w)}{1-\e^{-\i L
Y_L(w)}}\ \d w + \frac{1}{2\pi} \int_{C_2} \frac{K(w,z)Y'_L(w)}{
\e^{\i L Y_L(w)}-1}\ \d w\nn
&=& \frac{\i}{2\pi L}\int_0^\infty \frac{1}{1+e^x}\left[
K\Bigl(\xi+\frac{\i\, x}{LY_L'(\xi)},z\Bigr)-
K\Bigl(\xi-\frac{\i\, x}{LY_L'(\xi)},z\Bigr)
\right] \d x\nn
&& \mbox{} - (\xi \rightarrow \xi^*).
\label{eq:fscorr1}
\eea
If the endpoints $\xi,\xi^*$ are such that we can Taylor-expand the
kernels appearing in \r{eq:fscorr1}, we can simplify the expression
further with the result
\bea
{\cal A} &\approx& - \frac{K'(\xi,z)}{\pi L^2 Y_L'(\xi)}\int_0^\infty \frac{x}{1+e^x}\ \d x\ -\ (\xi \rightarrow \xi^*)\nn
&=& -\frac{\pi}{12L^2}\frac{K'(\xi,z)}{Y_L'(\xi)}\ -\ (\xi \rightarrow \xi^*).
\label{eq:fscorr2}
\eea
This is the leading Euler-MacLaurin correction term that occurs in
the low and high density phases, see \r{eq:intY_M}. The key in the
above derivation was the ability to expand
\bea
K\Bigl(\xi+\frac{\i\, x}{LY_L'(\xi)},z\Bigr)&-&
K\Bigl(\xi-\frac{\i\, x}{LY_L'(\xi)},z\Bigr)
\nn
&&\sim
\ln\left[\frac{LY_L'(\xi)\bigl(z^{-1}-\xi\bigr)+\i\, x}
{LY_L'(\xi)\bigl(z^{-1}-\xi\bigr)-\i\, x}\right]
\eea
in a power series in $x$. This is unproblematic as long as
$LY_L'(\xi)\bigl(z^{-1}-\xi\bigr)$ is large, which turns out to be the
case in the low and high density phases as well as on the coexistence
line.  

In the maximum current phase the above analysis no longer holds. As
mentioned in the main text, $\xi\to -1$ in the MC phase, and this has
two consequences. Firstly, by virtue of the endpoints $\xi,\xi^*$
being close to stationary point, the derivative $Y_L'(\xi)$ is now
found to be of order $L^{-1/2}$ whereas $Y_L''(\xi)$ is of order
one. As a result (c.f. \cite{BAring2}) we have to retain
subleading terms in the Taylor-expansions of $Y_L(w)$ and $Y'_L(w)$
around $w=\xi$, e.g.  
\be
Y_L(w)\sim Y_L(\xi)+Y_L'(\xi) (w-\xi)+
\frac{1}{2}Y_L''(\xi)(w-\xi)^2.
\label{eq:Yexp}
\ee

Secondly, and more seriously, as $\xi-\xi^{-1} = \mathcal{O}(L^{1/2})$ the Taylor
expansion of the integration kernel $K(\xi,z)$ breaks down near
$z=\xi$. Taking \r{eq:Yexp} into account and following through the
same steps as before, we find that the leading contributions of the
last two terms in \r{eq:intYapp} in the maximum current phase are 
\bea
{\cal A}&=&\frac{\i}{\pi L}\int_0^\infty\frac{x+p}{1+\exp(x^2+2xp)}
K\Bigl(\xi+\varepsilon x,z\Bigr)\nn
&&-\frac{\i}{\pi L}\int_0^\infty\frac{x-\i p}{1+\exp(x^2-2\i p x)}
K\Bigl(\xi+\i\, \varepsilon x,z\Bigr)\ -\ (\xi\rightarrow\xi^*).
%&& \mbox{}- \frac{\i}{\pi L}\int_0^\infty\frac{x+z_2}{1+\exp(x^2+2xz_2)}
%K\Bigl(\xi^*+\varepsilon_2 x,z\Bigr)\nn
%&&+\frac{\i}{\pi L}\int_0^\infty\frac{x+\i z_2}{1+\exp(x^2+2\i xz_2)}
%K\Bigl(\xi^*-\i\, \varepsilon_2 x,z\Bigr).
\eea
Here we have introduced
\be
\varepsilon=\sqrt{\frac{2\i}{L Y_L''(\xi)}}=\mathcal{O}(L^{-1/2}),\qquad
%\varepsilon_2=\sqrt{\frac{2\i}{L Y_L''(\xi^*)}},\\
p = -\frac{\i}{2} \varepsilon Y'_L(\xi) L =
\mathcal{O}(1).
%,\qquad z_2 = -\frac{\i}{2} \varepsilon_2 Y'_L(\xi^*) L = \mathcal{O}(1). 
\ee
In order to make progress we would like to expand around
$x=0$. However, as $z^{-1}-\xi$ becomes of order ${\cal  O}(L^{-1/2})$
for $z$ near $\xi$, all the terms $x^n K^{(n)}(\xi,z)$ in the
expansion are of the same
order $L^{0}$ for ($n\geq 1$), and we would need to self-consistently
determine the full series. The $n=0$ terms in the Taylor
expansion, which are proportional to $\ln(L)L^{-1}$, cancel, and
we find that ${\cal A}={\cal O}(L^{-1})$.

%%%%%%%%%%%%%%%%%%%%%%%%%%%%%%%%%%%%%%%%%%%%%%%%%%
\section{Expansion coefficients\label{ap:Mcoef}}
%%%%%%%%%%%%%%%%%%%%%%%%%%%%%%%%%%%%%%%%%%%%%%%%%%

In the following list we abbreviate $y'_n(z_{\rm c})$ by $y'_n$,
\bea
\kappa_1 &=& - y_0' \delta_1,\\
\lambda_1 &=&  y_0' \frac{\eta_1}{\pi},\\
\kappa_2 &=& -y_0' \delta_2 - y_1' \delta_1 -
  \frac12 y_0'' (\delta_1^2 - \eta_1^2), \\ 
\lambda_2 &=& \frac1{\pi}\left( y'_0 \eta_2 + y'_1\eta_1 + y_0'' \delta_1
\eta_1\right),\\ 
\mu_2 &=& y'_0 \frac{\delta_1\eta_1}{\pi},\\
\kappa_3 &=& -y_0'\delta_3 - y_1'\delta_2 -y_2'\delta_1 - y_0'' (\delta_1\delta_2 -
\eta_1\eta_2) -\frac12 y_1''(\delta_1^2-\eta_1^2) \nonumber\\
&& {} - \frac16 y_0'''\delta_1(\delta_1^2 - 3\eta_1^2), \\
\lambda_3 &=& \frac1\pi \left( y_0'\eta_3 + y_1'\eta_2 + y_2'\eta_1 +
y_0''(\delta_1\eta_2 +\delta_2\eta_1) + \delta_1\eta_1y_1''\right.
\nonumber\\
&&\left. {} +\frac16 y_0''' \eta_1 (3\delta_1^2 - \eta_1^2) \right),\\
\mu_3 &=& \frac1\pi \left( y_0'(\delta_1\eta_2 + \delta_2\eta_1) +
y_1'\delta_1\eta_1 + \frac13 y_0''\eta_1(3\delta_1^2-\eta_1^2)
\right. \nonumber\\
&& \left. {} +
\frac{\pi^2}{3} \frac{y_0''}{y_0'^2} \eta_1\right),\\
\nu_3 &=& \frac1{6\pi}\left( y_0'\eta_1(3\delta_1^2-\eta_1^2) - 2\pi^2 \frac{\eta_1}{y'_0}\right).
\eea

%%%%%%%%%%%%%%%%%%%%%%%%%%%%%%%%%%%%%%%%%%%%%%%%%%%%%%%%%%
\section{Complex/Real excited states and their Bethe root
  distributions} 
\label{app:finestruc}
\begin{figure}[ht]
\begin{center}
\epsfxsize=0.5\textwidth
\epsfbox{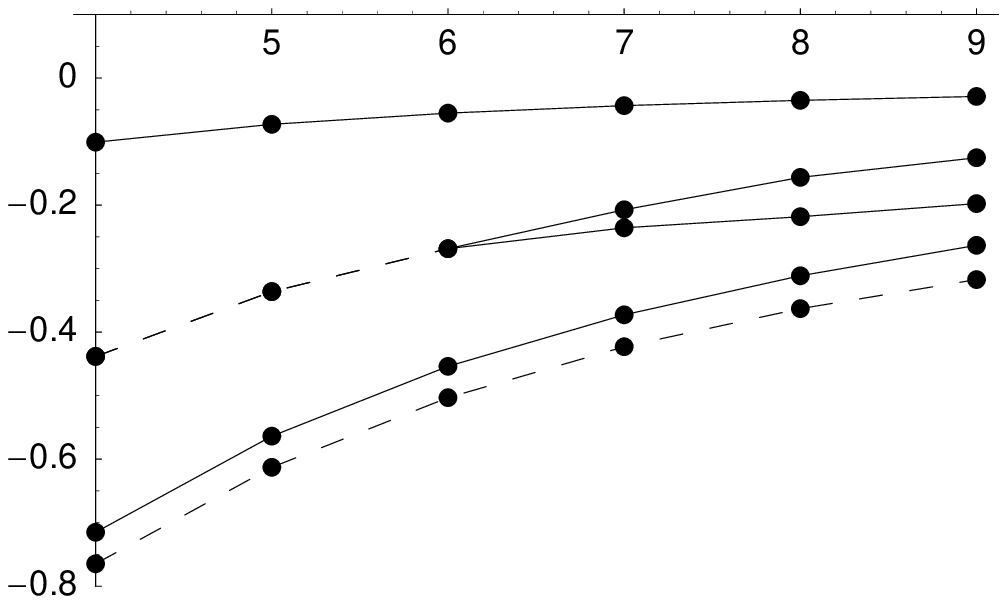}
\caption{Real parts of the eigenvalues of low-lying excited
  states for small system sizes $L=4,5,\ldots,9$. The lines are a guide to
  the eye only. Dashed lines indicate complex conjugate pairs of
  eigenvalues, solid lines correspond to 
  real eigenavlues.} 
\label{fig:excspec}
\end{center}
\end{figure}

%%%%%%%%%%%%%%%%%%%%%%%%%%%%%%%%%%%%%%%%%%%%%%%%%%%%%%%%%%
As mentioned in the main text, excited states can change from complex-
to real-valued with increasing 
lattice length. In order to illustrate this point we consider the
TASEP with $\alpha=\beta=0.25$. In Fig. \ref{fig:excspec} we plot the
real parts of the eigenvalues of the transition matrix for lattice
lengths $L=4,5,\ldots,9$, computed by direct diagonalisation. The
lowest excitation always corresponds to a real eigenvalue. For
$L=4,5,6$ the excited states with second largest real part are a pair
of complex conjugated eigenvalues. For larger $L$ however, this pair
becomes a pair of real eigenvalues. In terms of the Bethe roots this
comes about in the following way. The pair of complex conjugated
eigenvalues corresponds to distributions of integers 
\be
I_j = -L/2 +j \quad\mathrm{for}\quad j=1,\ldots,L-2\qquad I_{L-1} =
L/2,
\label{int:s1}
\ee
and
\be
I_j = -L/2 +j \quad\mathrm{for}\quad j=2,\ldots,L-1\qquad I_{1} = -L/2,
\label{int:s2}
\ee
respectively. The numerical values of the corresponding Bethe roots
for even $L$ are listed in Table \ref{tab:exc}. When the lattice
length exceeds $L=6$ the nature of the excited states changes. Now
there are two {\sl real} eigenvalues corresponding to the distribution
\r{int:s2} of integers, see Table \ref{tab:exc} for the $L=8$ site
system. Analogous violations of the one-to-one correspondence between sets of integers
and roots of the Bethe ansatz equations have been previously observed
in both ``ferromagnetic'' \cite{finestruc} and ``antiferromagnetic'' 
\cite{EFS} situations.

\begin{table}
\begin{center}
\begin{tabular}{|l|l|l|l|}
\hline
$L$ & $E$ & integer & root\\ \hline \hline
$4$ & $-0.438276 + 0.118338 \i$ & $\ \ 2$  & $z_1=  -0.538337 + 0.536475\i$\\
    &                        & $\ \ 0$  & $z_2=   0.212018 + 0.064141\i$\\
    &                        & $-1$ & $z_3=   0.199931 - 0.122095\i$\\ \hline
$4$ & $-0.438276 - 0.118338\i$ & $-2$  & $z_1=  -0.538337 - 0.536475\i$\\
    &                        & $\ \ 0$  & $z_2=   0.212018 - 0.064141\i$\\
    &                        & $\ \ 1$ & $z_3=   0.199931 + 0.122095\i$\\ \hline
$6$ & $-0.268648 + 0.037915\i$ & $\ \ 3$ & $z_1=  -0.737950 + 0.284318\i$\\
    &                        & $\ \ 1$ & $z_2=   0.145219 + 0.179042\i$\\
    &                        & $\ \ 0$ & $z_3=   0.193853 + 0.050586\i$\\
    &                        & $-1$  & $z_4=   0.190087 - 0.072453\i$\\
    &                        & $-2$  & $z_5=   0.130508 - 0.204632\i$\\ \hline
$6$ & $-0.268648 - 0.037915\i$ & $-3$ & $z_1=  -0.737950 - 0.284318\i$\\
    &                        & $-1$ & $z_2=   0.145219 - 0.179042\i$\\
    &                        & $\ \ 0$ & $z_3=   0.193853 - 0.050586\i$\\
    &                        & $\ \ 1$ & $z_4=   0.190087 + 0.072453\i$\\
    &                        & $\ \ 2$ & $z_5=   0.130508 + 0.204632\i$\\ \hline
$8$ & $-0.156354$             & $\ \ 4$ & $z_1=  -0.549451$\\
    &                        & $\ \ 2$ & $z_2=  0.062316 + 0.229156\i$\\
    &                        & $\ \ 1$ & $z_3=  0.141563 + 0.134657\i$\\
    &                        & $\ \ 0$ & $z_4=  0.173073 + 0.044259\i$\\
    &                        & $-1$   & $z_5=  0.173073 - 0.044259\i$\\
    &                        & $-2$   & $z_6=  0.141563 - 0.134657\i$\\
    &                        & $-3$   & $z_7=  0.062316 - 0.229156\i$\\ \hline
$8$ & $-0.218262$             & $\ \ 4$ & $z_1=  -1.20057$\\
    &                        & $\ \ 2$   &$z_2=  0.0876066 + 0.252770\i$\\
    &                        & $\ \ 1$   &$z_3=  0.163307 + 0.144776\i$ \\
    &                        & $\ \ 0$   &$z_4=  0.193253 + 0.0471657\i$\\
    &                        & $-1$   &$z_5=  0.193253 - 0.0471657\i$\\
    &                        & $-2$   &$z_6=  0.163307 - 0.144776\i$\\
    &                        & $-3$   &$z_7=  0.0876066 - 0.252770\i$\\
\hline
\end{tabular}
\caption{Low-lying excited states for $\alpha=\beta=0.25$ and $L=4,6,8$.}
\label{tab:exc}
\end{center}  
\end{table}
%%%%%%%%%%%%%%%%%%%%%%%%%%%%%%%%%%%%%%%%%%%%%%%%%%%
\section{Excited States of the TASEP on a Ring}
\label{app:ring}
%%%%%%%%%%%%%%%%%%%%%%%%%%%%%%%%%%%%%%%%%%%%%%%%%%%
In this appendix we collect some results on the excitation spectrum of
the TASEP on a ring, i.e. with periodic boundary conditions. Golinelli
and Mallick have recently developed a simple method for calculating
the spectral gap in the TASEP on a ring \cite{GoliM04}. Their method
applies to higher excited states as well. In what follows we consider
the case of half-filling only, i.e. $L/2$ particles on a ring
with $L$ sites.
The Bethe ansatz equations of the half-filled TASEP are \cite{BAring1,BAring2}
\be
(1-Z_j^2)^{L/2}=-2^L\prod_{k=1}^{L/2}\frac{Z_k-1}{Z_k+1}\
,\quad j=1,\ldots,\frac{L}{2}.
\label{eq:ringBAE}
\ee
\be
E=\sum_{j=1}^{L/2}\frac{Z_j-1}{2}\ .
\ee
As the right hand side of \r{eq:ringBAE} is independent of the index
$j$, we may write \cite{BAring1,BAring2,GoliM04}
\be
(1-Z_j^2)^{L/2}=-e^{\pi u}\ ,\quad -1\leq{\rm Im}(u)<1.
\label{eq:ringBAE2}
\ee
The $L$ roots of this equation are
\be
Z_m=-Z_{L/2+m}=\sqrt{1-y_m}\ ,\quad m=1,\ldots,\frac{L}{2},
\ee
where
\be
y_m=e^{2\pi(u+\i)/L}\ e^{4\pi \i(m-1) /L}.
\ee
The $y_m$'s lie on a circle with radius $e^{2\pi u/L}$ such that
\be
0\leq {\rm arg}(y_1)<{\rm arg}(y_2)<\ldots<{\rm arg}(y_{L/2})<2\pi\ .
\label{eq:args}
\ee
In order to construct a particular solution of the Bethe ansatz
equations, one chooses a sequence of roots $y_{c(1)},\ldots,
y_{c(L/2)}$ where $1\leq c(1)<\ldots<c(n)\leq L$
and then determines the parameter $u$
self-consistently from
\be
e^{\pi u}=2^L\prod_{k=1}^{L/2}\frac{Z_{c(k)}-1}{Z_{c(k)}+1}.
\label{eq:selfC}
\ee
%%%%%%%%%%%%%%%%%%%%%%%%%%
\subsection{Ground State}
\begin{figure}[ht]
\begin{center}
\epsfxsize=0.3\textwidth
\epsfbox{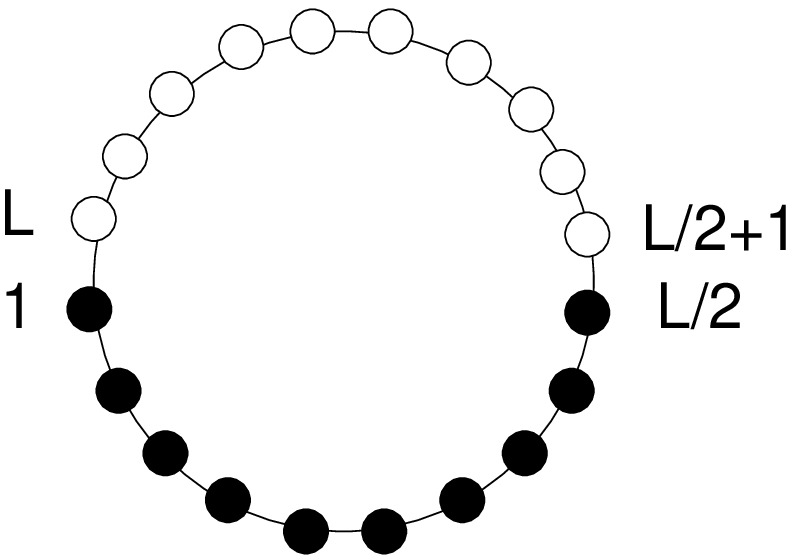}
\caption{Distribution of roots $y_m$ corresponding to the ground state
  of the TASEP on a ring.}
\label{fig:TASEPGS}
\end{center}
\end{figure}

%%%%%%%%%%%%%%%%%%%%%%%%%%
The ground state of the TASEP on a ring is obtained by choosing
$c(j)=j$, $j=1,\ldots, L/2$. The corresponding distribution of roots
$y_m$ is shown in Figure~\ref{fig:TASEPGS}.

%%%%%%%%%%%%%%%%%%%%%%%%%%%%%%%%%%%%%%%
\subsection{Particle-Hole Excitations}
%%%%%%%%%%%%%%%%%%%%%%%%%%%%%%%%%%%%%%%

\begin{figure}[ht]
\begin{center}
(a)\epsfxsize=0.3\textwidth
\epsfbox{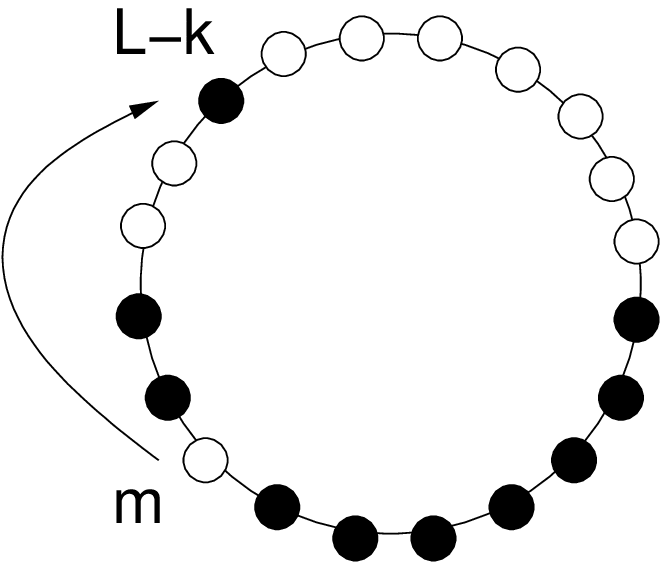}
\qquad
(b)\epsfxsize=0.3\textwidth
\epsfbox{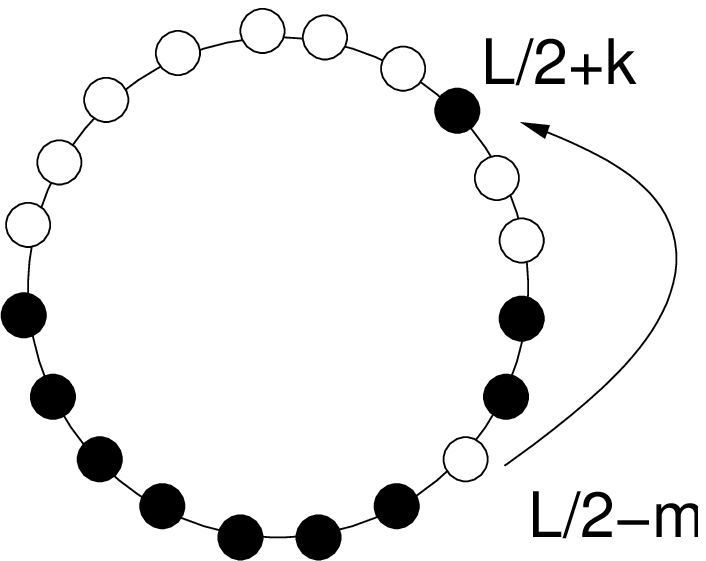}
\caption{Particle-hole excitations with energies (a) $E_{m,L-k}$ and
  (b) $E_{L/2-m,L/2+k}$.}
\label{fig:TASEPTI}
\end{center}
\end{figure}

Some simple excited states can be constructed by choosing sequences of
the type
\be
c(j)=\left\{\begin{array}{ll}
j & {\rm if}\ j<m\\
j+1 & {\rm if}\ m\leq j<\frac{L}{2}\\
L-k & {\rm if}\ j=\frac{L}{2}.
\end{array}
\right.
\label{eq:sequence}
\ee
Such excitations correspond to having a hole at $y_m$ and an extra
particle at $y_{L-k}$ as compared to the ground state. The roots
distribution is shown in Figure~\ref{fig:TASEPTI}(a). In what follows
we are interested in the limit $L\to\infty$, while keeping $k$ and $m$
fixed. 

Golinelli and Mallick \cite{GoliM04} have shown how to simplify and
solve the self-consistency condition \r{eq:selfC} in the large-$L$
limit for the first excited state, which corresponds to the choice
$(m,k)=(1,0)$ in \r{eq:sequence}. Following through exactly the same
steps for general $(m,k)$, we obtain the following self-consistency
condition for the parameter $u$ in the limit $L\to\infty$
\be
{\rm Li}_{3/2}\bigl(-e^{\pi u}\bigr)+2\pi
\left[\sqrt{-u-(2m-1)\i}+\sqrt{-u+(2k+1)\i}\right]=0.
\label{eq:u}
\ee
Once $u$ is known from a numerical solution of \r{eq:u} the leading
behaviour of the eigenvalue of the transition matrix can be determined from
\be
E_{m,L-k}=a_{m,L-k}(u)\left[\frac{2}{L}\right]^{3/2}+\ldots
\ee
\bea
a_{m,L-k}(u)&=&
\frac{\pi^{3/2}}{6}\left[\bigl(-u-(2m-1)\i\bigr)^{3/2}
+\bigl(-u+(2k+1)\i\bigr)^{3/2}\right]\nn
&&
-\frac{1}{8\sqrt{\pi}}{\rm Li}_{5/2}\bigl(-e^{\pi u}\bigr).
\eea
The momentum of the particle-hole excitations is given by
\be
P_{m,L-k}=-(m+k)\frac{2\pi}{L}.
\ee
We have solved \r{eq:u} for the first few excited states of this type
and list the corresponding values of $u$ and the coefficients
$a_{m,L-k}$ that characterize the asymptotic behaviour eigenvalues of
the transition matrix in Table \ref{tab:ring}.

\begin{table}
\begin{center}
\begin{tabular}{|l|l|l|}
\hline
$(m,k)$  & $u$ 	   &$a_{m,L-k}$\\ \hline\hline
$(1,0)$ & $	1.11907$ &	 $-6.50919$\\ \hline
$(2,0)$ & $	1.55661 - 0.18053 \i$ &	 $-17.1884 - 5.43662 \i$\\ \hline
$(1,1)$ & $	1.55661 + 0.18053 \i$ &	 $-17.1884 + 5.43662 \i$\\ \hline
$(2,1)$ & $	1.90545$             &	 $-28.9435$\\ \hline
$(1,2)$ & $	1.84895 + 0.313567 \i$ &	 $-30.6202 + 14.0523 \i$\\ \hline
$(3,0)$ & $	1.84895 - 0.313567 \i$ &	 $-30.6202 - 14.0523 \i$\\ \hline
$(2,2)$ & $	2.15672 + 0.141419 \i$ &	 $-43.0859 + 8.57338 \i$\\ \hline
$(3,1)$ & $	2.15672 - 0.141419 \i$ &	 $-43.0859 - 8.57338 \i$\\ \hline
$(1,3)$ & $	2.07629 + 0.422166 \i$ &	 $-46.2912 + 25.1677 \i$\\ \hline
$(4,0)$ & $	2.07629 - 0.422166 \i$ &	 $-46.2912 - 25.1677 \i$\\ \hline
\end{tabular}
\caption{Eigenvalues of some low-lying excited states for the
  half-filled TASEP on a ring.}
\label{tab:ring}
\end{center}
\end{table}
The choices $(m,k)=(2,0)$ and $(m,k)=(1,1)$ give the eigenvalues with
maximum real part and non-zero imaginary parts.

A second sequence of simple particle-hole excitations is obtained by
the choice of sequence
\be
c(j)=\left\{\begin{array}{ll}
j & {\rm if}\ j<\frac{L}{2}-m\\
j+1 & {\rm if}\ \frac{L}{2}-m\leq j<\frac{L}{2}\\
\frac{L}{2}+k & {\rm if}\ j=\frac{L}{2}.
\end{array}
\right.
\label{eq:sequence2}
\ee
This corresponds to having a hole at $y_{L/2-m}$ and an extra
particle at $y_{L/2+k}$. It is easy to see that energy and
momentum of such excitations are given by
\be
E_{L/2-m,L/2+k}\equiv E_{k,L-m},
\ee
\be
P_{L/2-m,L/2+k}=(m+k)\frac{2\pi}{L}.
\ee

%%%%%%%%%%%%%%%%%%%%%%%%%%%%%%%%%%%%%%%%%%%%%%%%%
\subsection{Multiple Particle-Hole Excitations}
\begin{figure}[ht]
\begin{center}
\label{fig:TASEP4p}
(a)\epsfxsize=0.3\textwidth
\epsfbox{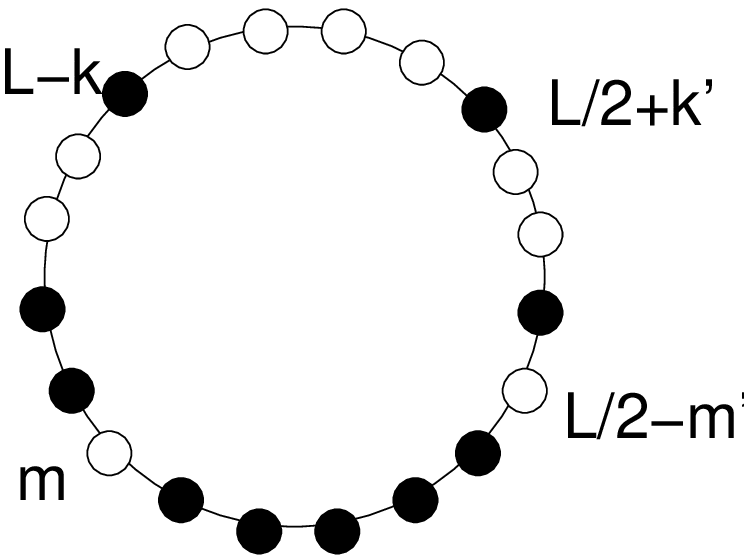}
(b)\epsfxsize=0.25\textwidth
\epsfbox{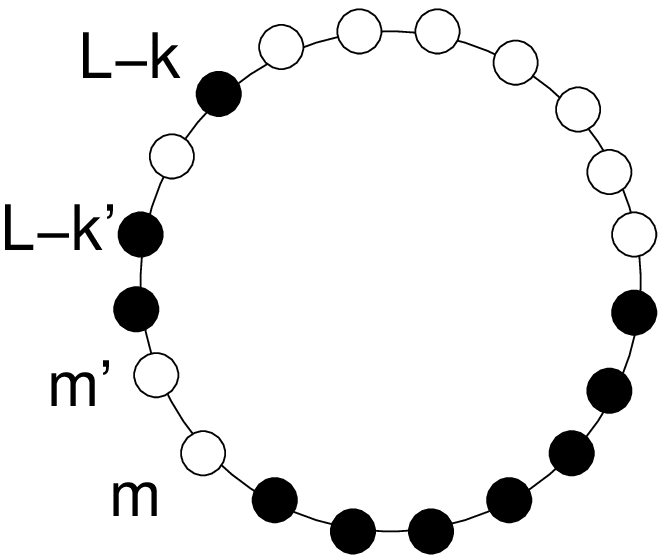}
(c)\epsfxsize=0.28\textwidth
\epsfbox{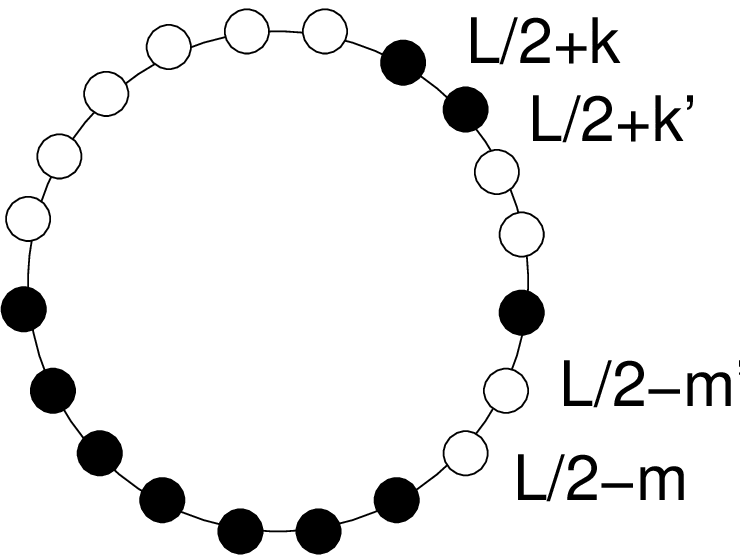}
\caption{Two-particle two-hole excitations.}
\end{center}
\end{figure}

%%%%%%%%%%%%%%%%%%%%%%%%%%%%%%%%%%%%%%%%%%%%%%%%%
Multiple particle-hole excitations can be constructed along the same
lines. We note that while the momentum is additive, i.e. simple the
sum of the momenta of the constituent particle-hole excitations, this
is not the case for the energy. Hence the Bethe-ansatz particle and
holes still interact with one another. Let us first consider a
two-particle two-hole excitation characterized by the sequence

\be
\renewcommand{\arraystretch}{1.2}
c(j)=\left\{\begin{array}{ll}
j & {\rm if}\ j<m\\
j+1 & {\rm if}\ m\leq j<\frac{L}{2}-m\\
j+2 & {\rm if}\ \frac{L}{2}-m\leq j<\frac{L}{2}-1\\
\frac{L}{2}+k' & {\rm if}\ j=\frac{L}{2}-1\\
L-k & {\rm if}\ j=\frac{L}{2}.
\end{array}
\right.
\label{eq:sequence3}
\ee
This corresponds to having two holes at positions $y_m$ and
$y_{L/2-m'}$ and two extra particles at 
$y_{L/2+k'}$ and $y_{L-k}$, see Figure~\ref{fig:TASEP4p}(a).

Following once more the procedure of \cite{GoliM04} we arrive at the
following equation determining the parameter $u$ in the $L\to\infty$ limit
\bea
&&{\rm Li}_{3/2}\bigl(-e^{\pi u}\bigr)+2\pi
\Bigl[\sqrt{-u-(2m-1)\i}+\sqrt{-u+(2k+1)\i}\nn
&&\qquad+\sqrt{-u+(2m'+1)\i}+\sqrt{-u-(2k'-1)\i}\Bigr]=0.
\label{eq:u4p}
\eea
The transition matrix eigenvalue of the two-particle two-hole
excitation for very large $L$ is given by 
\be
E_{m,L-k,L/2-m',L/2+k'}=a_{m,L-k,L/2-m',L/2+k'}(u)
\left[\frac{2}{L}\right]^{3/2}+\ldots
\ee
\bea
&&a_{m,L-k,L/2-m',L/2+k'}(u)=
\frac{\pi^{3/2}}{6}\Bigl\{\bigl(-u-(2m-1)\i\bigr)^{3/2}\nn
&&\quad+\bigl(-u+(2k+1)\i\bigr)^{3/2}
+\bigl(-u+(2m'+1)\i\bigr)^{3/2}\nn
&&\quad+\bigl(-u-(2k'-1)\i\bigr)^{3/2}\Bigr\}
-\frac{1}{8\sqrt{\pi}}{\rm Li}_{5/2}\bigl(-e^{\pi u}\bigr).
\label{eq:a4p}
\eea
The momentum is equal to
\be
P_{m,L-k,L/2-m',L/2+k'}=(m'+k'-m-k)\frac{2\pi}{L}.
\ee
We have solved \r{eq:u4p} and \r{eq:a4p} for some low-lying excited
states and list the results in Table \ref{tab:ring2}.

\begin{table}
\begin{center}
\begin{tabular}{|l|l|l|}
\hline
$(m,k,m',k')$    & $u$                           &
$a_{m,L-k,\frac{L}{2}-m',\frac{L}{2}+k'}(u)$\\
\hline\hline
$ (1,0,0,1) $	 & $1.65874$	                 & $-16.0176$\\ \hline
$ (1,1,0,1) $	 & $1.99633 + 0.124903 \i$	 & $-27.8453 + 5.01165 \i$\\ \hline
$ (2,0,0,1) $	 & $1.99633 - 0.124903 \i$	 & $-27.8453 - 5.01165 \i$\\ \hline
$ (1,0,1,1) $	 & $1.99633 + 0.124903 \i$	 & $-27.8453 + 5.01165
\i$\\ \hline
$ (1,0,0,2) $	 & $1.99633 - 0.124903 \i$	 & $-27.8453 - 5.01165
\i$\\ \hline
$ (2,0,0,2) $	 & $2.31089 - 0.205544 \i$	 & $-40.3126 - 9.52328 \i$\\ \hline
$ (1,1,1,1) $	 & $2.31089 + 0.205544 \i$	 & $-40.3126 + 9.52328 \i$\\ \hline
$ (2,0,1,1) $	 & $2.29142$	                 & $-40.4698$\\ \hline
$ (1,1,0,2) $	 & $2.29142$	                 & $-40.4698$\\ \hline
$ (1,0,1,2) $	 & $2.29142$	                 & $-40.4698$\\ \hline
$ (2,1,0,1) $	 & $2.29142$	                 & $-40.4698$\\ \hline
$ (1,2,0,1) $	 & $2.23876 + 0.239858 \i$	 & $-42.0218 + 13.3219 \i$\\ \hline
$ (1,0,2,1) $	 & $2.23876 + 0.239858 \i$	 & $-42.0218 + 13.3219 \i$\\ \hline
$ (2,1,1,1) $	 & $2.57296 + 0.0881572 \i$	 & $-53.6251 + 4.5669 \i$\\ \hline
$ (1,1,1,2) $	 & $2.57296 + 0.0881572 \i$	 & $-53.6251 + 4.5669 \i$\\ \hline
$ (2,0,1,2) $	 & $2.57296 - 0.0881572 \i$	 & $-53.6251 - 4.5669 \i$\\ \hline
$ (2,1,0,2) $	 & $2.57296 - 0.0881572 \i$	 & $-53.6251 - 4.5669 \i$\\ \hline
\end{tabular}
\caption{Eigenvalues of some low-lying two-particle two-hole excited states for the
  half-filled TASEP on a ring.}
\label{tab:ring2}
\end{center}
\end{table}
The two-particle two-hole excitations shown in Figures~\ref{fig:TASEP4p}(b) and (c) can be
constructed along the same 
lines. The transition matrix eigenvalues in the large-$L$ limit fulfil
\bea
E_{m,L-k,m',L-k'}&=&E_{m,L-k,L/2-k',L/2+m'}\ ,\nn
E_{L/2-m,L/2+k,L/2-m',L/2+k'}&=&
E_{k,L-m,L/2-m',L/2+k'}\ .
\eea

\end{document}